\documentclass[a4paper,fleqn,usenatbib]{mnras}

 \usepackage{newtxtext,newtxmath}  

\usepackage[T1]{fontenc}
\usepackage{ae,aecompl}

\usepackage{graphicx}			
\usepackage{amsmath}			
\usepackage{amssymb}		
\usepackage{mathtools}  		
\usepackage{cite}
\usepackage{natbib}
\usepackage{hhline}
\usepackage[normalem]{ulem}  
\usepackage{booktabs}   		

\newcommand{\HA}{H$\alpha$}	   				 
\newcommand{\MS}{M$_\odot$}	    				 
\newcommand{\angstrom}{\textup{\AA}}		

\DeclareRobustCommand{\VAN}[3]{#3}    

\title[Near-identical SFRD (\text{H}$\alpha$ \& FUV)  at redshift zero]{Near-identical star formation rate densities from \text{H}$\alpha$ and FUV at redshift zero}

\author[F.M. Audcent-Ross et al.]{
\noindent Fiona M. Audcent-Ross,$^{1}$\thanks{E-mail: fiona.audcent-ross@icrar.org}
Gerhardt R. Meurer,$^{1}$
O.I. Wong,$^{1,2}$
Z. Zheng,$^{3}$
\newauthor
D. Hanish,$^{4}$
M.A. Zwaan,$^{5}$
J. Bland-Hawthorn,$^{6}$
A. Elagali,$^{1}$
M. Meyer,$^{1,2}$
\newauthor
M.E. Putman,$^{7}$
E.V. Ryan-Weber,$^{2,8}$
S.M. Sweet,$^{8}$
D.A. Thilker,$^{4}$
\newauthor
M. Seibert,$^{9}$
R. Allen,$^{10}$
M.A. Dopita,$^{11}$
M.T. Doyle-Pegg, $^{12}$
M. Drinkwater, $^{12}$
\newauthor
H.C. Ferguson,$^{10}$
K.C. Freeman,$^{11}$
T.M. Heckman,$^{4}$
R.C. Kennicutt, Jr.,$^{13}$
\newauthor
V.A. Kilborn,$^{8}$
J.H. Kim,$^{14}$
P.M. Knezek,$^{15}$
B. Koribalski,$^{16}$
R.C. Smith, $^{17}$
\newauthor
L. Staveley-Smith,$^{1}$
R.L. Webster$^{18}$ and
J.K. Werk$^{19}$
\\
\textit{$^{1 } $ International Centre for Radio Astronomy Research, University of Western Australia, 35 Stirling Highway, Crawley, WA 6009, Australia} \\
$^{2 } $ ARC Centre of Excellence for All-Sky Astrophysics (CAASTRO), Australia\\
$^{3 } $ National Astronomical Observatories, Chinese Academy of Sciences, 20A Datun Rd, Beijing 100012, China\\
$^{4 }$ Department of Physics and Astronomy,  John Hopkins University, 3400 North Charles Street, Baltimore, MD 21218, USA\\  
$^{5 }$ European Southern Observatory, Karl-Schwarzschild-Strasse 2, D-85748 Garching bei M\"{u}nchen, Germany \\
$^{6 }$ Sydney Institute for Astronomy, University of Sydney, Sydney, NSW 2006, Australia \\
$^{7 }$ Department of Astronomy, Columbia University, 550 West 120th Street, New York, 10027, USA  \\
$^{8 }$ Centre for Astrophysics and Supercomputing, Swinburne University of Technology, Hawthorn, VIC 3122, Australia \\  
$^{9}$ Observatories of the Carnegie Institution for Science, Pasadena, CA 91101, USA \\
$^{10}$ Space Telescope Science Institute, 3700 San Martin Drive, Baltimore, MD 21218, USA \\
$^{11 }$ Research School of Astronomy and Astrophysics (RSAA), Australian National University, Cotter Road, Weston Creek, ACT 2611, Australia \\
$^{12}$ Department of Physics, University of Queensland, Brisbane, QLD 4072, Australia \\
$^{13}$ Institute of Astronomy, University of Cambridge, Madingley Road, Cambridge CB3 0HA, UK  \\
$^{14}$ Subaru Telescope, National Astronomical Observatory of Japan, 650 North A'ohoku Place, Hilo, HI 96720, USA\\
$^{15}$ National Science Foundation, 4201 Wilson Boulevard, Arlington, Virginia 22230, USA     \\
$^{16}$ Australia Telescope National Facility, CSIRO, P.O. Box 76, Epping, NSW 1710, Australia  \\
$^{17}$ Cerro Tololo Inter-American Observatory (CTIO), Casilla 603, La Serena, Chile  \\
$^{18}$ School of Physics, University of Melbourne, VIC 3010, Australia   \\
$^{19}$ Astronomy Department, University of Washington, 3910 15th Ave. NE, Seattle, WA 98195-0002, USA   \\
}
\date{Accepted xx. Received YYY; in original form ZZZ}

\pubyear{2015}

\begin{document}
\label{firstpage}
\pagerange{\pageref{firstpage}--\pageref{lastpage}}
\maketitle

\begin{abstract}
For the first time both \HA{} and far-ultraviolet (FUV) observations from an \textsc{Hi}-selected sample are used to determine the dust-corrected star formation rate density (SFRD: $\dot{\rho}$) in the local Universe.  Applying the two star formation rate indicators on 294 local galaxies we determine log($\dot{\rho} _{\textnormal{H}\alpha}) = -1.68~^{+0.13}_{-0.05}$ ~[M$_{\odot} $ yr$^{-1} $ Mpc$^{-3}]$ and log($\dot{\rho}_{\textnormal{FUV}}$) $ = -1.71~^{+0.12}_{-0.13}$ [M$_\odot $ yr$^{-1} $ Mpc$^{-3}]$.  These values are derived from scaling \HA{} and FUV observations to the \textsc{Hi} mass function.   
Galaxies were selected to uniformly sample the full \textsc{Hi} mass (M$_{\text{\textsc{Hi}}}$) range of the \textsc{Hi} Parkes All-Sky Survey (M$_{\text{\textsc{Hi}}} \sim10^{7}$ to $\sim10^{10.7}$ \MS).  The approach leads to relatively larger sampling of dwarf galaxies compared to optically-selected surveys.  The low \textsc{Hi} mass, low luminosity and low surface brightness galaxy populations have, on average, lower \HA{}/FUV flux ratios than the remaining galaxy populations, consistent with the earlier results of Meurer.  The near-identical \HA- and FUV-derived SFRD values arise with the low \HA{}/FUV flux ratios of some galaxies being offset by enhanced \HA{} from the brightest and high mass galaxy populations.  Our findings confirm the necessity to fully sample the \textsc{Hi} mass range for a complete census of local star formation to include lower stellar mass galaxies which dominate the local Universe.  
\smallskip

\end{abstract}




\section{Introduction}

The star formation rate density (SFRD: $\dot{\rho}$) of the local Universe provides an important observational constraint on cosmological theories explaining the formation and evolution of galaxies and, therefore, on the build-up of stellar mass since the Big Bang.  By combining ultraviolet (UV), optical, infrared and radio continuum survey results, \citet{RN86} and \citet{RN416} showed how SFRD varies with redshift.  In the subsequent two decades there has been considerable research quantifying the evolution of $\dot{\rho}$  \citep[for a summary see] []{RN81}.  There is a growing consensus that the SFRD of the Universe peaked at  \mbox{$z\sim 1.9$, $\sim3.5$ Gyr} after the Big Bang and then declined exponentially to the current epoch  \citep[e.g., see][]{RN465, RN325, RN332, RN81}.

Different star formation tracers can be used to measure the local SFRD, and fluxes from the \HA{} emission line and the far-ultraviolet (FUV) continuum are commonly used.  Each tracer has its own strengths and biases \citep[see the overview in][]{RN81}.  \HA{} provides a direct estimate of the ionising output of a stellar population, and thus its content of ionising O-type stars.  As such it provides a direct measure of recent massive star formation and does not require adjustment for factors such as chemical abundances, unlike other emission line tracers \citep[e.g.,][]{RN638}.  Flux calibration, active galactic nuclei (AGN) contamination, stellar absorption, initial mass function (IMF) selection and dust extinction need to be considered, however, for \HA{} surveys making SFRD measurements. Prominent and recent \HA{} surveys include \citet{RN465, RN644, RN648, RN645, RN328,RN483}.     See \citet{RN328} for a useful compilation of SFRD measurements derived from narrowband surveys.

The ultraviolet continuum ($\lambda  \sim 912 - 3000$ \AA) is dominated  by the emission of O- and B-type stars \citep[][]{RN322} and thus is sensitive to the formation of somewhat lower mass stars than \HA{} emission, and hence of longer main sequence lifetimes.  With the advent of the GALEX satellite most of the sky has been imaged in the near and far ultraviolet  \citep{RN769}.  FUV-derived SFRD measurements require sizeable corrections for flux attenuation by dust \citep[e.g.,][]{RN639, RN640}, with considerable spread ($\sim 1$ mag for z $\sim$ 0) in the estimates made for this important correction \citep{RN81}.  Widely cited and recent UV-derived SFRD measurements  include \citet{RN636,RN583,RN642,RN643, RN679} and see the compilation in \citet[][]{RN81}.

The selection of the sample used to estimate the SFRD of the local Universe is also important in making an accurate measurement \citep{RN41}.  Ideally all galaxies in a large volume of the local Universe should have their star formation rate (SFR) measured.  Many surveys use optically-selected samples, although such surveys have  well known biases against low luminosity and low surface brightness (LSB) galaxies \citep[e.g.,][]{RN593, RN323}.  \textsc{Hi}-selection provides an alternative method for choosing the input sample for SFRD studies.  It avoids the biases of optical selection and ensures the sample has an interstellar medium (ISM), a necessary condition for star formation \citep[e.g.,][]{RN70}.  While star formation occurs in a molecular medium \citep[e.g.][]{RN672, RN676,RN674}, molecular ISM has proven difficult to detect in low luminosity and LSB galaxies, while \textsc{Hi}\textbf{ } is readily found \citep{RN677, RN320,RN674,RN673, RN483}. An \textsc{Hi}-selected sample, therefore, helps to give a wide range of local gas-rich, star-forming galaxies but excludes gas-poor galaxies which typically have negligible star formation, such as  early-types and dwarf spheroids  \citep[e.g.,][]{RN41, RN674, RN588}.   \textsc{Hi}-selection also tends to disfavour high density environments such as galaxy clusters (which also typically show little star formation), while favouring low density filaments and voids \citep{RN660, RN258}. \citet{RN18} and \citet{RN483} have previously calculated the local SFRD using \HA{} observations on \textsc{Hi}-selected samples.  

Until recent decades there have been very few galaxy surveys utilising two independent SFR tracers on a homogeneous sample \citep[][]{RN585,RN648, RN689,RN592}.  Those with rigorously-selected samples provide an invaluable way to examine and directly calibrate the differences  between the two SFR measurements, including at both extremes of the luminosity functions \citep[e.g.,][]{RN649,RN583,RN34,RN596}. 

For the first time we report on both \HA{} and FUV observations of an \textsc{Hi}-selected sample of galaxies, thereby enabling a direct comparison of the SFRD ($z \sim 0$) values arising from these two commonly-used SFR indicators in the local Universe.  

Targets for the Survey of Ionization in Neutral Gas Galaxies \citep[SINGG;][] {RN41} and the Survey of Ultraviolet emission of Neutral Gas Galaxies \citep[SUNGG;][]{RN381} were chosen to thoroughly sample the \textsc{Hi}  properties of galaxies.   The same number of targets in each decade of \textsc{Hi} mass (M$_{\text{\textsc{Hi}}}$) were selected, to the extent allowed by the parent sample, with the nearest targets at each \textsc{Hi} mass chosen for observation.  The data typically contain just one \textsc{Hi}  source per set of multiwavelength images.  This approach allows reasonable sampling of the full range of the \textsc{Hi} mass function (HIMF) with limited telescope resources. It also allows us to derive volume densities by scaling to the HIMF, using the method employed by \citet{RN18}.

The paper is organised as follows: Section \ref{sec:Our data_sec} outlines the two surveys, SFR calibrations, sample selection and the HIMF-based methodology we use to determine the SFRD for the local Universe.  Section \ref{sec:Results_sec} presents the results of our  calculations and details the systematic differences observed in \HA{}/FUV flux ratios.  Section \ref{sec:Discussion_sec} shows how near-identical SFRD values arise despite the systematic differences between the two SFR indicators.  We present our conclusions in Section \ref{sec:Conclusions_sec}.   

The \citet{RN317} single power-law IMF over a mass range of 0.1 -- 100 \MS, a Hubble constant of $H_0= $ 70 ~km s$^{-1}$ Mpc$^{-1}$ and cosmological parameters for a $\Lambda$CDM cosmology of $\Omega_{0}=0.3$ and $\Omega_{\Lambda} =0.7$ have been used throughout this paper.

\section{Data and methodology}
\label{sec:Our data_sec}

\begin{table*}
	\centering
   
	\begin{tabular}{lllllcll}
	& & \\
	   HIMF comparison \\    
		\hline
		
	     HIMF & $      \alpha$ & log M$_{*}$ & $\theta_{*}$ & log $\rho_{\text{\textsc{Hi}}}$ & log($\dot{\rho}{_{{\textnormal{H}\alpha}}}$)  & log($\dot{\rho}_{\textnormal{FUV}}$)  & \textsc{Hi} Survey \\
	     (1) & (2) & (3) & (4) & (5) &  (6) &   (7) \\
	     \hline
This work: \\

 \citet{RN318}  & $-1.37 \pm 0.06$ & $9.86 \pm 0.04$ & $4.9 \pm 1.0$  & $ 7.71$ & $ -1.68~^{+0.13}_{-0.05}$  & $ -1.71~^{+0.12}_{-0.13}$   &HIPASS\\  
 \\
Other HIMFs:		 & &  \\
\citet{RN18} & $-1.41 \pm 0.05$ & $9.92 \pm 0.04$ & $3.9 \pm 0.7$ & $ 7.70$  & $ -1.68~^{+0.13}_{-0.05}$ & $ -1.71~^{+0.12}_{-0.14}$ & selected from HIPASS \\ [0.075cm] 
 
\citet{RN51}  & $-1.24$ & $9.99$ & $3.2$  & $7.61 $ & $-1.75~^{+0.14}_{-0.05} $  & $-1.78~^{+0.14}_{-0.17}$ & see \citet{RN52} \\  [0.075cm]

\citet{RN39}  & $-1.33 \pm 0.02$ & $9.96 \pm 0.02$ & $4.8  \pm 0.3$ & $7.79 $  & $-1.58~^{+0.13}_{-0.05}$ & $-1.61~^{+0.13}_{-0.16}$  & ALFALFA ($\sim$10k sample)\\  [0.075cm] 

\citet{RN22}  & $-1.37 \pm 0.03$ & $10.06 \pm 0.04 $ & $2.65 \pm 0.5$ & $7.66 $ &$-1.70~^{+0.14}_{-0.05} $ & $ -1.72~^{+0.14}_{-0.16}$ & AUDS (60$\%$ complete)\\ [0.075cm] 

\citet{RN678}& $-1.25 \pm 0.04$ & $9.94 \pm 0.04$ & $4.5 \pm 0.4$ & $7.70$  & $-1.65~^{+0.14}_{-0.05} $  & $-1.69~^{+0.14}_{-0.16}$  & ALFALFA (final) \\

		 & &  \\
		\hline
			 & &  \\
	
	\end{tabular}
		\caption{\textsc{Hi} mass density ($\rho_{\text{\textsc{Hi}}}$) and dust-corrected SFRD results using the listed HIMF models and SFR calibration Equations \ref{eqn: SFRHA_eqn} (\HA) and \ref{eqn: SFRFUV_eqn} (FUV).  Column descriptions [units]: Col. (1): Source reference. Col. (2): Schechter fit power-law slope.  Col. (3): Schechter fit characteristic \textsc{Hi} mass [\MS]. Col. (4): Schechter fit normalisation [$ \times 10^{-3}$ Mpc$^{-3}$ dex$^{-1}$].   Col. (5): \textsc{Hi} mass density [\MS{} Mpc$^{-3}$], calculated using the listed HIMF (Col. 1) and Eqn. \ref{eqn: SFRHA_eqn}.    Cols. (6 -- 7):  SFRD derived from \HA{} and FUV observations, respectively, using the named HIMF and Equations \ref{eqn: SFRHA_eqn} and \ref{eqn: SFRFUV_eqn}, respectively.   Cols. (2 -- 7) Random and systematic errors have been added in quadrature, where applicable.  See Section \ref{subsection:HIMFdiscussion_sec} for further discussion.
 \label{table:HIMF_table}}

\end{table*}


\subsection{SINGG Survey}
\label{subsection:SINGG_sec}

SINGG samples galaxies from the \textsc{Hi} Parkes All-Sky Survey \citep [HIPASS:][]{RN321, RN319, RN320}.  \citet{RN18} sets out the approach taken here to calculate the SFRD in detail, and the \citet{RN318}  HIMF parameters used are listed in Table~\ref{table:HIMF_table}.  SINGG observations were made with both \textit{R}-band and narrowband \HA{} filters to isolate \HA{}.  \HA{} emission (at rest $\lambda$ = 6562.82 \AA) primarily arises as a result of  the photoionisation of H\textsc{ii} regions around high mass (M$_{*}  \gtrsim  ~20$ \MS{}), short-lived ($t < 10$ Myr) O-type stars.

The processing used on SINGG's first data release \citep{RN41, RN18} has been applied to the SINGG sample of 466 galaxies from 288 HIPASS objects (see Meurer 2018, in prep.).  The distances and corrections for [\textsc{Nii}] contamination, stellar absorption, and foreground and internal dust absorption  are unchanged from \citet{RN41}.  Optical observations are corrected for internal dust attenuation in accordance with the empirical relationship of \citet{RN584}, using uncorrected \textit{R}-band absolute magnitudes and Balmer line ratios \citep[see][]{RN41}.     

To ensure all star-forming areas were identified for each HIPASS target, an examination of the SINGG three-colour FITS images  was undertaken (primarily by FAR and GM).  Apertures were set in a consistent manner, ensuring all detectable \HA{} emission from the targets was included.

\subsection{SUNGG Survey}
\label{subsect:SUNGG_sec}

SINGG's sister survey, SUNGG, measured NUV (2273 \AA) and FUV (1515 \AA) fluxes.  UV emission arises from both O- and B-type stars  and consequently traces a wider range of initial masses \linebreak
(M$_{*} \gtrsim 3$ \MS{}) and stellar ages than \HA{} emission. 

SUNGG observed 418 galaxies from 262 HIPASS objects at both FUV and NUV wavelengths \citep[][]{RN381, RN257}.  We use FUV as our SFR tracer as it is not as contaminated by hot old stellar remnants (white dwarfs) as the NUV band is   \citep[e.g.,][]{RN681,RN583,RN582}.

The SUNGG survey processing used in this work is largely unchanged from \citet{RN322} and \citet{RN381}, and will be described in Wong 2018 (in prep.).  SUNGG corrects for foreground galactic extinction using the reddening maps from \citet{RN428} and applying the \citet{RN427} extinction law.  The FUV correction for internal dust attenuation is  unchanged from \citet[][]{RN257}, and is based on the FUV-NUV colour and utilises the low redshift algorithm of \citet{RN583}.


\begin{figure}
\centering
\includegraphics[scale=0.41]{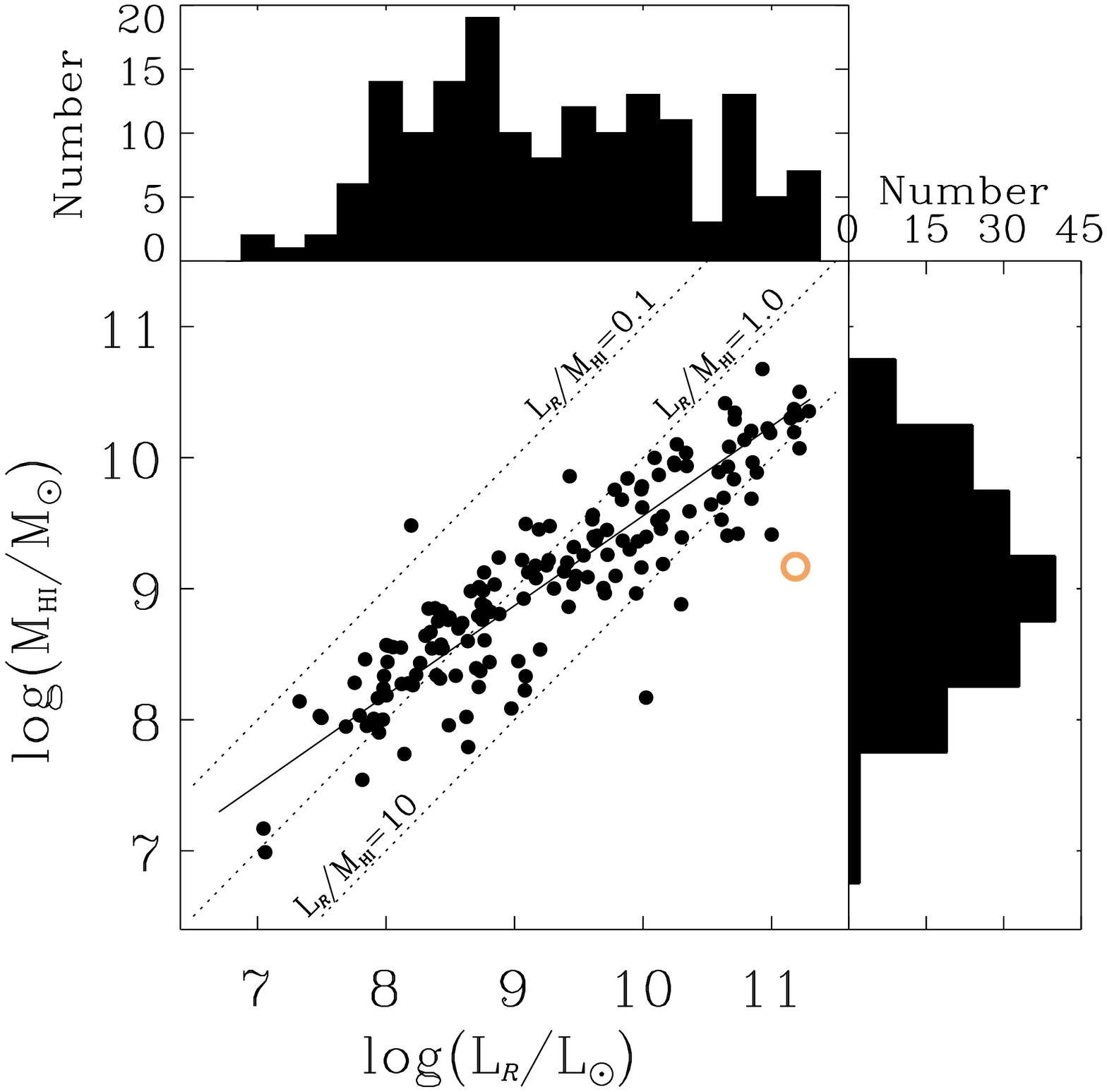}
\caption{\textsc{Hi} mass (M$_{\text{\textsc{Hi}}}$) plotted against \textit{R}-band luminosity (L$_{R}$/L$_\odot$) for the 160 single galaxies in the sample, with the ordinary least squares fit (Y vs. X) to the data shown as a solid line. L$_{R}$ is a crude proxy for stellar mass and the three dashed lines reflect the constant luminosity-to-\textsc{Hi} mass fractions of 0.1, 1.0 and 10, as marked.  The median error on  log(L$_R$/L$_\odot$) is smaller than the symbol used.  J0242+00  is overlaid with a large open orange circle (see Appendix \ref{sect:rem_gal_section} for further discussion on this galaxy).  The outlier J1247-03 (log(L$_{R}$/L$_\odot) = 10.0 $ and log(M$_{\text{\textsc{Hi}}}/$M$_\odot) = 8.17$) is discussed in Section \ref{subsec:Luminositydensities_sec} and Table \ref{table:outliers_table}.  \label{fig:massdistplot }}  
 \end{figure}

\subsection{SFR calibrations}
\label{subsect:SFRcalcs_eqn}

The  \HA-derived SFR (SFR$_{\textnormal{H}\alpha}$) for each SINGG galaxy is calculated assuming solar metallicity and continuous star formation, and applies a \citet{RN317} single power-law IMF over the birth mass range of 0.1 to 100 \MS, which we adopt throughout.  The \citet{RN322} $\textnormal{SFR}_{\textnormal{H}\alpha}{}$ calibration is applied and compared to the \citet{RN329} calibration (in parentheses):  

\begin{equation}
\label{eqn: SFRHA_eqn}
\indent
\textnormal{SFR}_{\textnormal{H}\alpha}{}~ [\textnormal{M}_{\odot}~ $yr$^{-1}] = \frac{\textnormal{L}_{\textnormal{H}\alpha} ~\textnormal{[ergs s}^{-1}]}{1.04~(1.27) ~\times ~10^{41}~}.                   
\end{equation}

\smallskip

The FUV-derived star formation rate (SFR$_{\textnormal{FUV}}$) is calculated using the \citet{RN322} SFR$_{\textnormal{FUV}}$ calibration, with the \citet{RN329} calibration in parentheses,: 

\begin{equation}
\label{eqn: SFRFUV_eqn}
\indent
\textnormal{SFR}_{\textnormal{FUV}} ~[\textnormal{M}_{\odot}~ $yr$^{-1}] = \frac{\textnormal{L}_{\textnormal{FUV}} ~\textnormal{[ergs ~ s}^{-1}~\angstrom^{-1}]}{9.12~(9.09)~ \times ~10^{39}}.  
\end{equation}  

\smallskip

\citet{RN322} and \citet{RN329} star formation calibrations are derived with identical assumptions on the IMF slope and mass limits but the calibrations use different stellar populations models: Starburst99 \citep{RN601} and \citet{RN602}, respectively.

\subsection{The sample}
\label{subsect:Samples used here_sec}

The combined SINGG/SUNGG sample analysed here comprises the 294 galaxies that have flux measurements in four bands: \textit{R}, \HA{}, NUV and FUV.  Two galaxies (J0145-43 and J1206-22) meeting the above criteria are not included in the final sample, due to severe foreground star contamination.     

One further galaxy, J0242+00 (NGC 1068), is shown in several figures but is excluded from the final SFRD calculations.  It is remarkably luminous for its \textsc{Hi} mass and would increase $\dot{\rho} _{\textnormal{H}\alpha}$ and  $\dot{\rho} _{\textnormal{FUV}}$ by 36 and 13 per cent, respectively, if it was included in the sample.  Appendix \ref{sect:rem_gal_section} discusses the galaxy and the disproportionate effects it would have on our survey, if it was incorporated into the sample.    

HIPASS provides the total \textsc{Hi} mass of the target, with no ability to distinguish individual galaxies within the 15' beam of the Parkes 64-metre telescope.  The 294 galaxies analysed in this paper arise from 210 HIPASS targets.  Of these targets, 160 are single galaxies and the remaining 50 are systems with two or more galaxies, containing a total of 134 galaxies.  For  \textsc{Hi} sources comprised of multiple galaxies, we sum the luminosities (\HA, FUV and \textit{R}-band) of the individual galaxies to get aggregate luminosities for the system.  

Eleven systems have one minor galaxy for which we have \HA{} data but not FUV data.  The \HA{} flux of each of these minor system members is at least an order of magnitude smaller than the flux of the most luminous galaxy in the system.  Despite the exclusion of the minor galaxy lacking FUV data, we assessed these systems as being materially complete and have, therefore, retained them in the sample.



\begin{figure*}
\centering
\includegraphics[scale=1.15]{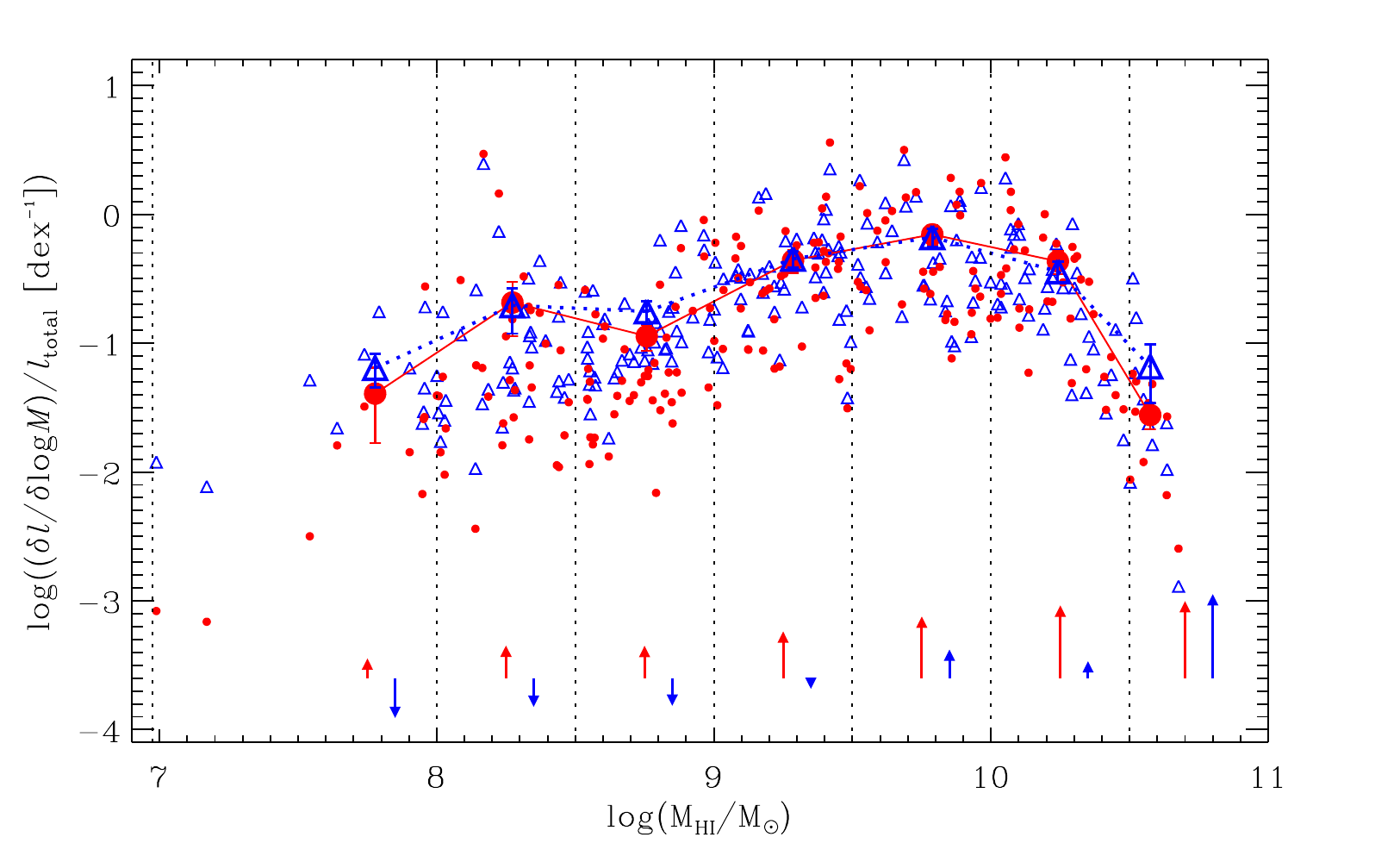}
\caption{Fraction of the total luminosity density per decade of \textsc{Hi} mass with \HA{} data shown in red and FUV in blue.  Mass bin limits are shown as vertical dashed lines. The small filled circles and small open triangles represent the contributions made by the 210 individual HIPASS targets to \HA{} and FUV luminosity densities, respectively.  The large symbols and associated error bars indicate the mean and $\pm1\sigma$ values for each log(M$_{\text{\textsc{Hi}}}/$M$_\odot$) bin's contribution, and the symbols are connected by lines (solid red for \HA{} and dotted blue for FUV) to guide the reader's eye.  All values are corrected for internal dust extinction.   The vectors at the bottom of each \textsc{Hi} mass bin illustrate the average change made to the fractional luminosity density from correction for internal dust extinction: \HA{} (left) and FUV (right).  
\label{fig:ldensitylist }}
\end{figure*}


After having excluded J0242+00, we make no further allowance for AGN contamination in the sample, as AGN are not likely to make a major  contribution to the total luminosity densities \citep[e.g.][]{RN648, RN687}.  Importantly, the impact of an AGN on the host's star formation activity lies within circumnuclear regions, which are typically dwarfed by the emission at larger radii \citep[e.g.,][]{RN40,RN464}. 

The SFRD values derived in this paper are local, with the 294 galaxies spanning distances of 3 to 135 Mpc, at an average value of $\sim$ 38 Mpc (median $\sim$ 20 Mpc).  This compares to the 110 galaxies in the first data release which, due to filter availability, were particularly local (median distance $\sim$ 13 Mpc) and were predominantly standalone, rather than group members.  The much larger sample used here spans  over 3.5 orders of magnitude in \textsc{Hi} mass and $\sim$4.5 dex in \textit{R}-band luminosity (see Fig. \ref{fig:massdistplot }).

\subsection{HIMF methodology}
\label{subsection:HIMF_methodology_sec}

In order to calculate volume-averaged quantities from a modest-sized sample, we scale our results to the HIMF and draw our sample from it as uniformly as possible. 

The \HA{} luminosity density, $l_{\textnormal{H}\alpha}$, for example, can be calculated using:
\begin{equation}
\label{eqn: HIMF_eqn}
\indent
l_{\textnormal{H}\alpha} = \int     \theta(M_{\text{\textsc{Hi}}}) \text{L}_{\text{H}\alpha}(M_{\text{\textsc{Hi}}}) d(M_{\text{\textsc{Hi}}}/M_{*})
\end{equation}

where $\theta(M_{\text{\textsc{Hi}}}) $ is the HIMF, the number density of galaxies as a function of \textsc{Hi} mass, \text{L}$_{\text{H}\alpha}$ is \HA{} luminosity and $M_{*}$ is the characteristic \textsc{Hi} mass of the Schechter parameterisation of the HIMF.  Following the binning of galaxies into {\textsc{Hi} mass bins, Eqn. \ref{eqn: HIMF_eqn} can be replaced with a summation (see \citet{RN18} Eqn. 3). \citet{RN18} explains the methodology of scaling our luminosity measurements to the HIMF in detail, together with the Monte Carlo and bootstrapping algorithms used to quantify the sampling and other random uncertainties from the approach.  Here, we use the HIMF from  \citet{RN318}.

The HIMF applied to the data is a source of possible systematic error in this method.  To determine the impact of the chosen HIMF, the SFRD and \textsc{Hi} mass density ($\rho_{\text{\textsc{Hi}}}$) calculations were repeated for each of the different HIMF options listed in Table \ref{table:HIMF_table}, keeping all other inputs unchanged.  The HIMFs tested include the recent HIMFs derived from the 60\% complete Arecibo Ultra-Deep Survey (AUDS) \citep{RN22}, the 40\% complete Arecibo Legacy Fast ALFA (ALFALFA) survey \citep{RN39}, the final ALFALFA catalog \citep{RN678} and  \citet{RN18}.  Utilising a HIPASS-selected sample, \citet{RN18}  obtains distances from \citet{RN605} and the \citet{RN424} model for deriving distances from radial velocities, allowing for infalling to nearby clusters and superclusters.  In contrast, the \citet{RN318} HIMF applied in this paper uses pure Hubble flow distances for the HIPASS survey.  See Sections  \ref{subsection:HIMFdiscussion_sec} and \ref{subsubsection{Distance_sec}}  for further discussion.

\section{Results}
\label{sec:Results_sec} 

\subsection{Luminosity densities and the local SFRD}
\label{subsec:Luminositydensities_sec} 

The \textit{R}-band, \HA{}, FUV and NUV luminosity density values derived from the sample are listed in Table~\ref{table:numberdensity_table}, with values given  before and after correction for internal dust.  Dust-corrected SFRD values $ \dot{\rho}_{{{\textnormal{H}\alpha}}}$
= 0.0211 and $\dot{\rho}_{{\textnormal{FUV}}}$ = 0.0197  [M$_\odot$ yr$^{-1} $Mpc$^{-3}$] are generated from Equations \ref{eqn: SFRHA_eqn} and \ref{eqn: SFRFUV_eqn}, respectively.  The quoted uncertainties correspond to an error of 11\% -- 35\%. 
The choice of SFR calibrations is a possible source of systematic error.  The \citet{RN322} calibrations and the widely adopted \citet{RN329} SFR calibrations (Equations \ref{eqn: SFRHA_eqn}  and \ref{eqn: SFRFUV_eqn}) were both applied, to aid comparisons with other studies.  Using \citet{RN329} generates values of log($\dot{\rho} _{\textnormal{H}\alpha})  = -1.76   $   and log($\dot{\rho}_{\textnormal{FUV}}) = -1.71$ [M$_{\odot} $yr$^{-1}  $ Mpc$^{-3}]$.


\begin{table*}
\begin{minipage} {\linewidth}
	\centering
\resizebox{0.80\linewidth}{!}{%
	\begin{tabular}{lllll} %
			 & &  \\

		Key values & \\
		\hline
		Quantity & Uncorrected  ~  &  Dust-corrected   ~ & Units & Notes\\
		
		\hline

		\textit{l$_{R}$} & ($4.8~^{+0.4}_{-0.5})  \times 10^{37}$ & ($7.6~^{+1.1}_{-0.9}) \times 10^{37}$ &   [ergs ~s$^{-1}$ Mpc$^{-3}$] & 1\\[0.075cm]

\textit{l}$_{\textnormal{H}\alpha}$ & ($ 9.5 ~^{+0.9}_{-1.0}) \times 10^{38}$ & ($2.2~^{+0.6}_{-0.3}) \times 10^{39}$ & [ergs ~s$^{-1}$ Mpc$^{-3}$]& 1 \\[0.075cm]   
		
		\textit{l}$_{\textnormal{FUV}}$ & ($4.8~^{+0.9}_{-1.2})  \times 10^{37}$ & ($1.8 \pm 0.5) \times 10^{38}$ &   [ergs \AA$^{-1}~$s$^{-1} $Mpc$^{-3}$]& 1\\ [0.075cm]
		
		\textit{l}$_{\textnormal{NUV}} $ &($2.9 ~\pm 0.2) \times 10^{37}$ & ($7.3 ~^{+2.0}_{-1.9}) \times 10^{37}$ &  [ergs \AA$^{-1}~$s$^{-1} $Mpc$^{-3}$] & 1\\  [0.075cm] 
&  & & \\
		 log($\dot{\rho}_{\textnormal{\HA}}$) & $-2.04 ~\pm$ 0.05 & $-1.68~^{+0.13}_{-0.05}$ & [M$_\odot ~$yr$^{-1} ~$Mpc$^{-3}]$ & 1   \\[0.075cm]
		 log($\dot{\rho}_{\textnormal{FUV}}$) & $-2.28~^{+0.12}_{-0.08}$  & $-1.71~^{+0.12}_{-0.13}$ & [M$_\odot ~$yr$^{-1} ~$Mpc$^{-3}]$  & 1  \\   
&  & & \\
		  $\rho_{\text{\textsc{Hi}}} $ & &  (5.2$~^{+1.0}_{-1.2})  \times 10^{7} $ & [M$_\odot ~$Mpc$^{-3}]$ \\  [0.075cm] 
$t_{\textnormal{gas}}$ (\HA)   & & $5.6 ~^{+1.9}_{-1.5} ~$ & [Gyr] & 2\\ [0.075cm] 
$t_{\textnormal{gas}}$ (FUV)   & & $6.0  ~^{{+2.1}}_{-1.6} ~$ & [Gyr] & 2\\ 

				\hline
					 & &  \\
	
	\end{tabular}}

\caption{Key derived values.  Notes: (1) Luminosity densities, calculated  using Equations \ref{eqn: SFRHA_eqn} and \ref{eqn: SFRFUV_eqn}, are shown before and after internal dust corrections.  \HA{} fluxes have also been corrected for [\textsc{Nii}] contamination. See Table \ref{table:errors_himfsimple_table} for more detailed analysis of the uncertainties.  (2) Both \HA-derived and FUV-derived volume-averaged gas cycling times ($ t_{\textnormal{gas}}  \approx ~2.3 \rho_{\text{\textsc{Hi}}} / \dot{\rho} $) are less than the Hubble time, consistent  with earlier findings \citep[e.g.,][]{RN28,RN18}.  
\label{table:numberdensity_table}}

	\end{minipage}
\end{table*}



\begin{figure*}
\centering
\includegraphics[scale=0.90]{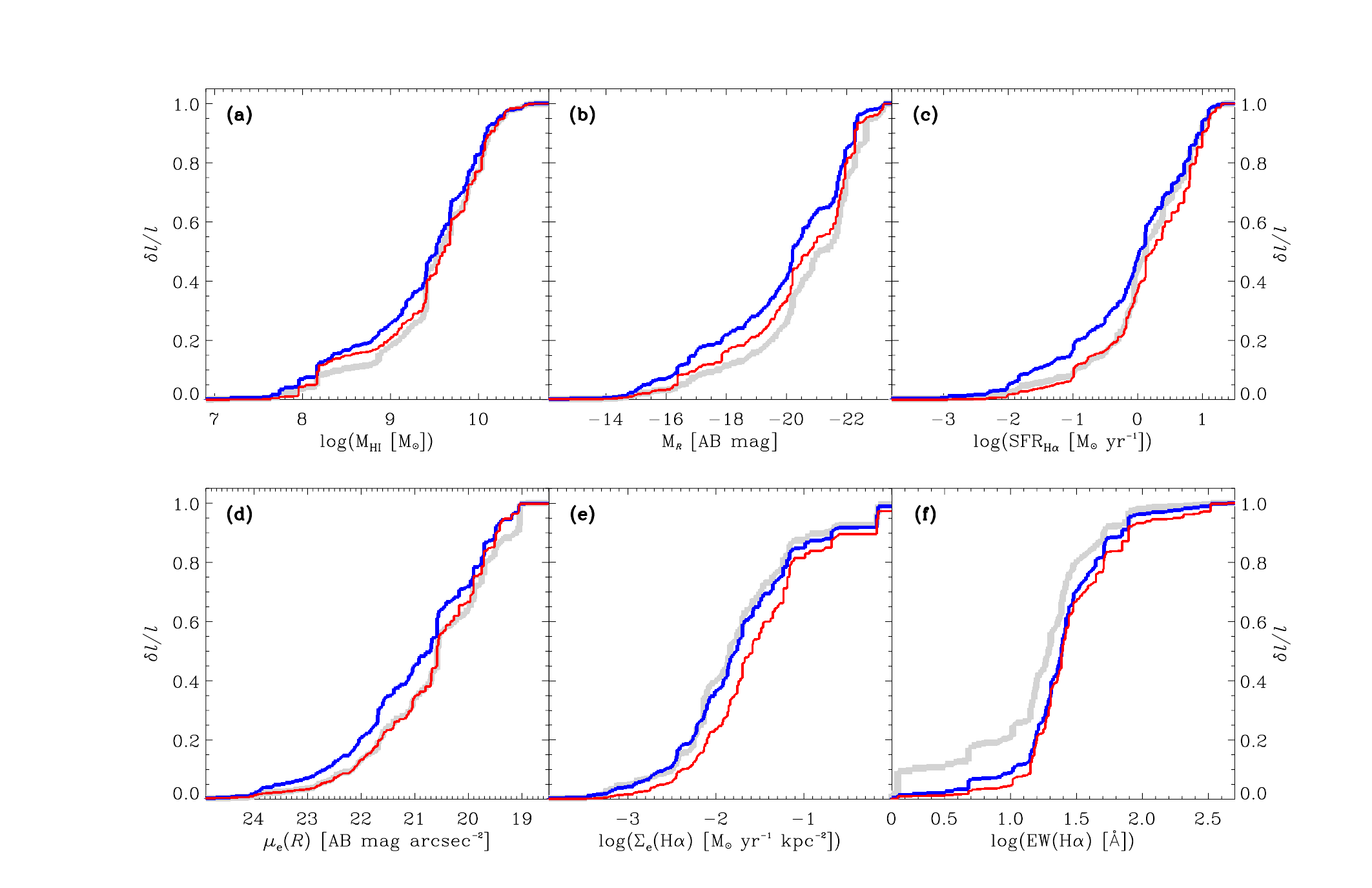}
\caption{Cumulative fractional luminosity densities (\textit{l}) analysed as a function of key quantities (a) \textsc{Hi} mass, (b) \textit{R}-band luminosity, (c) \HA{} star formation rate (see Eqn. \ref{eqn: SFRHA_eqn}), (d) \textit{R}-band effective surface brightness, (e) \HA{} surface brightness, and (f) \HA{} equivalent width (EW) (approximated by the ratio of \HA{} flux density to $R$-band flux density, consistent with \citet{RN18}).   Cumulative fractional luminosity densities shown are: \textit{R}-band (thick light grey line), \HA{} (thin red line), and FUV (thick dark blue line).  Plot (a): \textsc{Hi} mass: Note that the fluxes of the individual galaxies in multiple-galaxy targets have been totalled for this analysis of the 210 HIPASS targets (as described in Section \ref{subsect:Samples used here_sec}).  The low \textsc{Hi} mass targets make larger FUV fractional contributions than \HA, while larger \textsc{Hi}{} mass targets have higher \HA{} fractions. }
\label{fig:3fluxes_nohisto }
\end{figure*}

\begin{figure*}
\centering
{\includegraphics[scale=1.00]{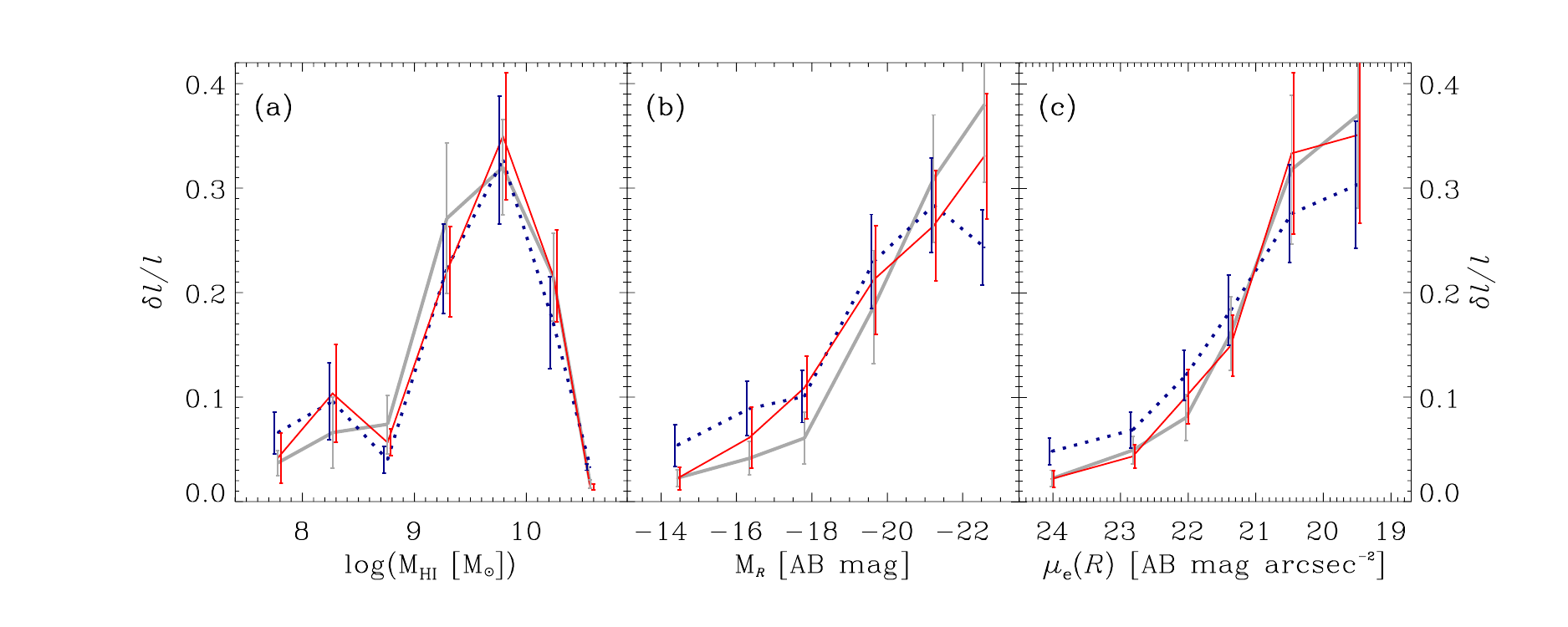}}  
\caption{Binned fractional luminosity densities (\textit{l}) for the parameters listed in Table \ref{table:ratios_full_table}: (a) log(M$_{\text{\textsc{Hi}}}$/\textnormal{M}$_{\odot}$) mass bins and (b -- c) binned data of $\sim$ 50 galaxies. The connecting lines are drawn to guide the reader's eye: \HA{} (thin red line), FUV (dotted blue line) and \textit{R}-band (thick grey line).  Error bars represent $\pm 1\sigma$, derived from 10,000 iterations of flux measurements adjusted by random, normally-distributed errors.  For clarity the error bars are slightly offset horizontally.} 
\label{fig:frac_lumindens }
\end{figure*}

The relative importance of each mass bin to the total luminosity density is shown in Fig. \ref{fig:ldensitylist }.  When comparing the contributions of different bins, note that the lowest mass bin is wider than the others, to ensure  all  bins contain a statistically significant number of galaxies.      Figure \ref{fig:ldensitylist } shows that the largest contribution to the total luminosity density is from the mass range log(M$_{\text{\textsc{Hi}}}$/\textnormal{M}$_{\odot}$) =  9.5 -- 10.5.  This bin includes the grand-design spiral galaxy J1338-17 (NGC 5247;  \citet{RN610}), the target with the largest impact on the SINGG/SUNGG~\textit{l}$_{\textnormal{H}\alpha}$ and \textit{l}$_{\textnormal{FUV}}$ values, comprising 4.6 and 3.9 per cent of the totals, respectively.  {} See Table \ref{table:outliers_table} for a list of galaxies with the highest impact on the total luminosity densities.

Individual galaxies within the two lowest \textnormal{\textsc{Hi}} mass bins also make significant contributions. J1247-03 (NGC 4691), for example, with a low \textsc{Hi} mass (log(M$_{\textnormal{\textsc{Hi}}}$/\MS) = 8.17), generates the second-highest \textit{l}$_{\textnormal{H}\alpha}$ and \textit{l}$_{\textnormal{FUV}}$ contributions (4.3  and 3.6 per cent, respectively). {} J1247-03 is a SBb peculiar galaxy with significant central star formation and supernovae activity (see \citet{RN609} for further discussion).   The lowest mass bin contributes the same, or more, per dex to the total \HA{} and FUV luminosity densities and SFRDs than the highest \textsc{Hi} mass bin (see columns (5) and (6) of Table \ref{table:ratios_full_table}\textit{a}).  Probing the low end of mass or luminosity functions is important.  \citet{RN581}, for example, increased their SFRD by $ \sim$ 0.07 dex  to compensate for incompleteness arising from faint galaxies in their optically-selected sample \citep[see also][]{RN328}.

\subsubsection{Cumulative fractional contributions}
\label{subsubsect:cumulfraccont_sec}

It is instructive to dissect how galaxies contribute to the SFRD as a function of key parameters.  We do this in Fig. \ref{fig:3fluxes_nohisto }, where we show the cumulative fractional contributions to \HA{}, FUV and \textit{R}-band luminosity densities (\textit{l}$_{\textnormal{H}\alpha}$, \textit{l}$_{\textnormal{FUV}}$, \textit{l}$_{R}$, respectively).  The \textit{R}-band flux from local galaxies originates primarily from established stellar populations and is, therefore, indicative of a galaxy's total stellar content. 

Figure \ref{fig:3fluxes_nohisto }\textit{a} illustrates the cumulative fractional contributions to the total \HA{}, FUV and \textit{R}-band luminosity densities as a function of \textsc{Hi} mass. Generally, targets in low \textsc{Hi} mass bins generate a higher fraction of the total \textit{l}$_{\textnormal{FUV}}$ compared to \textit{l}$_{\textnormal{H}\alpha}$ and \textit{l}$_{R}$.  Conversely, targets in \textsc{Hi} mass bins with log(M$_{\textnormal{\textsc{Hi}}}$/\MS) $\geqslant$ 10.0 have higher  $\textit{l}_{\textnormal{H}\alpha} $ fractional contributions (see also Fig. \ref{fig:frac_lumindens }\textit{a}).

Figures \ref{fig:3fluxes_nohisto }\textit{b} -- \ref{fig:3fluxes_nohisto }\textit{f} analyse the cumulative fractional luminosity densities for all 294 galaxies as a function of other key quantities.  Galaxies with low \textit{R}-band luminosity, low SFR$_{\textnormal{H}\alpha}$  values and LSB galaxies (both in \textit{R}-band and \HA) (Figs. \ref{fig:3fluxes_nohisto }\textit{b} -- \ref{fig:3fluxes_nohisto }\textit{e}), make lower fractional contributions to $\textit{l}_{\textnormal{H}\alpha}$ compared to $\textit{l}_{\textnormal{FUV}}$ (see also Figs. \ref{fig:frac_lumindens }\textit{b} -- \textit{c}).  Figures \ref{fig:3fluxes_nohisto }\textit{c} -- \ref{fig:3fluxes_nohisto }\textit{d} show that, for both SFR$_{\textnormal{H}\alpha}$ and \textit{R}-band surface brightness ($\mu_{R}$), $\textit{l}_{\textnormal{H}\alpha}$ follows $\textit{l}_{R}$, indicative of the total stellar content.     

Galaxies with little current star formation have low \HA{} equivalent width (EW) values (derived here from the SINGG $R$-band and \HA{} fluxes, consistent with \citet{RN18}) and, as expected, make low \HA{} and FUV fractional contributions, compared to the more dominant \textit{R}-band emission from their established stellar populations (Fig. \ref{fig:3fluxes_nohisto }\textit{f}).

\subsubsection{${\textnormal{H}\alpha}/$FUV ratios}
\label{subsubsectHAFUVratios_sec}   

The top panel of Table \ref{table:ratios_full_table} quantifies the fractional contributions made by the \textsc{Hi} mass binned data to the total luminosity density values, \textit{l}$_{\textnormal{H}\alpha}$, \textit{l}$_{\textnormal{FUV}}$ and \textit{l}$_{R}$.  The table highlights how the \textit{l}$_{\textnormal{H}\alpha}/ \textit{l}_{\textnormal{FUV}}$ ratios vary significantly across the ranges of \textsc{Hi} mass, \textit{R}-band luminosity  and $R$-band surface brightness.  The 50 galaxies with the faintest $R$-band surface brightness ($\mu_{R}$) have a small \textit{l}$_{\textnormal{H}\alpha}/ \textit{l}_{\textnormal{FUV}}$ ratio of 0.46 and contribute only 2.2 and 4.8 per cent  to the total \HA{} and FUV luminosity density values, respectively.  In contrast, the 44 galaxies with the brightest $\mu_{R}$ values contribute significantly to the \HA{} and  FUV luminosity densities (35  and 30 per cent, respectively) at a much higher $\textit{l}_{\textnormal{H}\alpha}/\textit{l}_{\textnormal{FUV}}$ ratio of 1.16.    See Section \ref{subsect:hafuvratio_sec} for further discussion.

The near-identical ${\textnormal{H}\alpha}$ and FUV SFRD values occur despite the differences noted above.  In particular, low surface brightness, low luminosity and low \textsc{Hi} mass galaxy populations make, on average, lower fractional contributions to $\textit{l}_{\textnormal{H}\alpha}$ than $\textit{l}_{\textnormal{FUV}}$, compared to the overall sample.


\begin{table*}
		Fractional luminosity densities analysed by key parameters

	\centering
  
\scriptsize{
 \resizebox{0.95\linewidth}{!}{%
	\begin{tabular}{llcllllll} 
		
		\hline
		 Parameter         &N   & Average   & \textit{l}$_{R}$ &\textit{l}$_{\textnormal{H}\alpha}$ & \textit{l}$_{\textnormal{FUV}}$ &  
		\textbf{}$\textit{l}_{\textnormal{H}\alpha}/\textit{l}_{{R}}$ & $\textit{l}_{\textnormal{H}\alpha}/\textit{l}_{\textnormal{FUV}}$  \\
		  Notes (1)&  (2)& values  (3) & (4) & (5) &  (6) & (7) & (8)\\   
		\hline
	     (a) log(M$_{\text{\textsc{Hi}}}$/\MS) \\
	     
	      6.975 -- 8.0 	& $11$ & 7.8	& 0.037 $  \pm $ 0.012 & 0.042 $ \pm $ 0.024 & 0.066 $ \pm $ 0.020  & 1.12 $ \pm $ 0.75   & 0.63 $ \pm  $ 0.42  \\
	      8.0 -- 8.5 & $34$& 8.3	& 0.066 $\pm $ 0.034  &  0.103 $ \pm$ 0.047  &  0.096 $ \pm $ 0.037 &   1.57 $ \pm $ 1.07  & 1.08 $ \pm $ 0.64  \\
	      8.5 -- 9.0 		& $42$ & 8.8	& 0.074 $ \pm $ 0.028 & 0.056 $ \pm $ 0.013 & 0.084 $ \pm $ 0.014 &  0.77 $ \pm$ 0.34  & 0.68 $ \pm $ 0.19    \\
	      9.0 -- 9.5	& $44$& 9.3  & 0.271 $ \pm $ 0.072 &  0.220 $  \pm $ 0.043 & 0.223 $ \pm $ 0.032 &  0.81 $ \pm $  0.27  & 0.98 $  \pm $ 0.24  \\
	      9.5 -- 10.0 & $34$& 9.8  & 0.320 $\pm $ 0.046 & 0.349 $ \pm $ 0.061 & 0.327 $ \pm $ 0.051 &  1.09 $ \pm $ 0.24 &  1.07 $ \pm $ 0.25  \\
	      10.0 -- 10.5	 & $35$& 10.2 & 0.215 $ \pm $ 0.042 & 0.216 $ \pm $ 0.044 & 0.171 $ \pm $ 0.031 &  1.01 $ \pm $ 0.29  &  1.26 $ \pm $ 0.35  \\
	      10.5 -- 11.0 	& $10$&10.6  &  0.017 $\pm $ 0.004 & 0.014 $ \pm $ 0.003 & 0.033 $ \pm $ 0.016 &  0.83 $ \pm $ 0.26   &  0.42 $ \pm $ 0.22 \\
	        
		& & & \\
		\hline
		& & & \\
	    (b) log(L$_R$ [L$_\odot$])  \\   
	    6.5  -- 8.1     & $50$ & 7.8   &  0.023 $ \pm $ 0.008 &  0.022 $ \pm $ 0.011 &  0.054 $ \pm $ 0.020  &  0.96 $ \pm $ 0.58  &  0.41 $ \pm $  0.25\\
	    8.1 -- 8.7      & $50$ & 8.5   & 0.041 $ \pm $ 0.016 &  0.061 $ \pm $ 0.029 &  0.089 $ \pm $ 0.026 &  1.49 $ \pm $ 0.92  &  0.69 $ \pm $ 0.38  \\
	    8.7  -- 9.4     & $50$ & 9.1 	& 0.061 $ \pm $ 0.025 &  0.109 $ \pm $ 0.030 &  0.101 $ \pm $ 0.025  &  1.79 $ \pm $ 0.88 &  1.08 $ \pm $   0.40     \\
	    9.4   -- 10.0  & $50$ &	9.8    & 0.186 $ \pm $ 0.054 &  0.213 $ \pm $ 0.052  &  0.230 $ \pm $ 0.045  &  1.15 $ \pm $ 0.43  &  0.93 $ \pm $ 0.29 \\
	    10.0  -- 10.6 & $50$ &10.4	& 0.309 $ \pm $ 0.061 &  0.264 $ \pm $ 0.053 &  0.283 $ \pm $ 0.045 &  0.85 $ \pm $ 0.24  &  0.93  $ \pm $  0.24 \\
	    10.6 -- 11.4  & 44     & 11.0	& 0.380 $ \pm $ 0.074 &  0.331 $ \pm $ 0.060 &  0.243 $ \pm $ 0.036 &   0.87 $ \pm $ 0.23   &  1.36 $ \pm $ 0.32 \\

		& & & \\
		\hline
		& & & \\
	     (c) $\mu_{R}$  \\
	     $[$AB mag arcsec$^{-2}$] \\
	     25.2 --  23.4 	& $50$  & 24.0 & 0.022  $ \pm $ 0.008 &  0.022  $ \pm $ 0.008 & 0.048 $ \pm $ 0.013  & 1.00 $ \pm $ 0.51     & 0.46 $ \pm $    0.21   \\
	     23.4 	--  22.4  & $50$  & 22.8 & 0.049  $ \pm $ 0.013 & 0.043  $ \pm $ 0.011  & 0.069 $ \pm $ 0.017 & 0.88 $ \pm $  0.32  & 0.62 $ \pm $ 0.22  \\
	     22.4 --  21.7  & $50$  & 22.0 &  0.080 $ \pm $ 0.022   &  0.101  $ \pm $ 0.026  &   0.121 $ \pm $ 0.024  &  1.26 $ \pm $ 0.48    & 0.84 $ \pm $ 0.27 \\
	     21.7  -- 21.0  & $50$  & 21.4 & 0.161 $ \pm $ 0.035 & 0.149  $ \pm $ 0.029  & 0.183 $ \pm $ 0.034   &  0.93 $ \pm $  0.27    & 0.81 $ \pm $ 0.22 \\
	     21.0  -- 20.0  & $50$  & 20.5 & 0.318  $ \pm $ 0.071  & 0.334  $ \pm $ 0.077  &  0.276 $ \pm $ 0.047  &  1.05 $ \pm $  0.34 &  1.21 $ \pm $  0.35\\
	     20.0  -- 17.5  & $44$  & 19.5  &  0.370  $ \pm $ 0.089  &  0.351  $ \pm $ 0.084  & 0.303 $ \pm $ 0.061 &  0.95 $ \pm $  0.32 & 1.16 $ \pm $ 0.36  \\

\hline
		\hline
			& & & \\
		& & & \\
	
	\end{tabular}}}
	\caption{Fractional luminosity density binned data for (a) the 210 targets analysed by \textsc{Hi} mass and the 294 galaxies analysed by (b) $R$-band luminosity and (c) \textit{R}-band surface brightness.  Notes: Col. (1) Bin limits for the listed parameters.  Col. (2) Number of targets (a) or galaxies (b and c). Col. (3) Average bin values for (a) \text{\textsc{Hi}} mass [M$_{\text{\textsc{Hi}}}$/\MS], (b) \textit{R}-band luminosity and (c) \textit{R}-band surface brightness.  Cols. (4 -- 6) Fractional contributions to the \textit{R}-band, \HA{} and FUV luminosity density values (\textit{l}$_{R}$, \textit{l}$_{\textnormal{H}\alpha}$, \textit{l}$_{\textnormal{FUV}}$), respectively.  Col. (7): Ratio of fractional contributions in \HA{} and \textit{R}-band, i.e., Cols. (5)/(4). Col. (8): Ratio of fractional contributions in \HA{} and FUV, i.e., Cols. (5)/(6). Cols. (4 -- 8) Quoted errors represent the standard deviation derived from 10,000 iterations of varying the underlying fluxes assuming normally-distributed errors.
\label{table:ratios_full_table}}
\end{table*}

\subsection{Star formation efficiency}
\label{subsection:resultsSFE_sec}


Star formation efficiency (SFE$_{\text{\textsc{Hi}}}$ = SFR/M$_{\text{\textsc{Hi}}}$) measures the star formation rate relative to the neutral hydrogen component of the ISM.  Although stars form from molecular gas, it is difficult to obtain molecular gas estimates, especially for low mass galaxies.  Hence SFE$_{\text{\textsc{Hi}}}$ remains a useful proxy measure of star formation  potential.  Figures \ref{fig:sixplot }\textit{a} --  \ref{fig:sixplot }\textit{c} show how SFE$_{\text{\textsc{Hi}}}$ varies as a function of key parameters for 129 of the single galaxies contained in the sample.   While SINGG  groups are not analysed in Fig. \ref{fig:sixplot }, \citet{RN323} showed that the larger SINGG groups had SFE$_{\text{\textsc{Hi}}}$ values consistent with the rest of the SINGG sample.        

The important differences between \HA{} and FUV fluxes noted in Section \ref{subsec:Luminositydensities_sec}, continue here, with low \textsc{Hi} mass, low \textit{R}-band luminosity and low surface brightness galaxies having systematically reduced SFE$_{\text{\textsc{Hi}}}$(\HA) compared to  SFE$_{\text{\textsc{Hi}}}$(FUV) (see Figs. \ref{fig:sixplot }\textit{a }-- \ref{fig:sixplot }\textit{c}).  

Log(SFE$_{\text{\textsc{Hi}}}$(FUV)) is little changed at $\sim -$9.8 \textnormal{yr$^{-1}$} over three decades of \textsc{Hi} mass (see Fig. \ref{fig:sixplot }\textit{a}), consistent with \citet{RN257}  (see also Table  \ref{table:sfrds_and_sfes_table}).  SFE$_{\text{\textsc{Hi}}}$(\HA), however, increases  by  $\sim$0.6~dex over the same range. The \HA{} best fit line has a slope of $0.21 \pm 0.05$ (see Table \ref{table:bestfits_table}), representing  a $\sim4 \sigma$ detection).

Galaxies  with low L$_{R}$ have systematically reduced SFE$_{\text{\textsc{Hi}}}$ values (Fig. \ref{fig:sixplot }\textit{b}), consistent with \citet{RN34}.  The trend is fractionally stronger in \HA{}, with a $\sim$1.3~dex variation in SFE$_{\text{\textsc{Hi}}}$(\HA) across the range of L$_{R}$.  Increasing SFE$_{\text{\textsc{Hi}}}$ with increasing \textit{R}-band surface brightness (Fig. \ref{fig:sixplot }$c$) mirrors the results of \citet{RN322}, \citet{RN330} and \citet{RN257}.   SFE$_{\text{\textsc{Hi}}}$(FUV) values increase $\sim 1.1$ dex and  SFE$_{\text{\textsc{Hi}}}$(\HA) by $\sim1.7$ dex over $\sim$ 6 orders of magnitude (both are $> 10 \sigma$ detections: see Table \ref{table:bestfits_table}).  

The gas cycling time-scale ($t_{\textnormal{gas}}$) is an estimate of the time taken for a galaxy to process its existing neutral and molecular ISM.  For consistency with \citet{RN41} we use the typical ISM H$_{2}/$\textsc{Hi} ratio determined by \citet{RN59}, M$_{\textnormal{gas}} = 2.3 $ M$_{\textnormal{\textsc{Hi}}}$, which gives $t_{\textnormal{gas}} (\textnormal{H}\alpha)  \approx ~2.3$ (M$_{\text{\textsc{Hi}}}/$ SFR$_{\textnormal{H}\alpha}$).  It is inversely proportional to SFE$_{\text{\textsc{Hi}}}$, therefore, as shown on the right hand axes of {Figs. \ref{fig:sixplot } {$a$ -- $c$}}.


\begin{table*}

SFRD and SFE$_{\textsc{Hi}}$ as a function of {\textsc{Hi}} mass \\     
\centering
\resizebox{0.95\linewidth}{!}{%
\begin{tabular}{lllllll} 
\hline
log(M$_{\text{\textsc{Hi}}}/$M$_\odot)  $ 	& N &	SFE$_{\text{\textsc{Hi}}} (\textnormal{H}\alpha) $	&	$\delta(\dot{\rho}_{\textnormal{H}\alpha}$) per 	&	$  l_{R}$ per    & SFE$_{\text{\textsc{Hi}}}$ (FUV) & $ \delta( \dot{\rho}_{\textnormal{FUV}}$) per    \\ 

& & $\times ~10^{-9 }$ & log(M$_{\text{\textsc{Hi}}}$/M$_\odot)$ bin  & log(M$_{\text{\textsc{Hi}}}/$M$_\odot)$ bin & $\times ~10^{-9 }$ & log(M$_{\text{\textsc{Hi}}}$/M$_\odot)$ bin\\  

 & &  & $\times ~10^{-3 }$ &$\times 10^{36}$ &  &  $\times ~10^{-3 }$ \\ 

(1) & (2) & (3) & (4) & (5) & (6) & (7) \\   
 \hline

$6.975-8.0$	&	$11$ &	 0.23 $\pm$  0.03 &	 0.86 $ \pm $  0.50  	 &  2.75  $ \pm $  0.90 & 0.34 $\pm$ 0.00 & 1.26 $\pm$ 0.38  \\  

$8.0-8.5$	&	$34$	&  $0.56 \pm 0.05$	 &  4.35 $\pm$  1.97 &  10.0  $\pm$  5.2 &  0.47 $\pm$ 0.01 &  3.79 $\pm$  1.45  \\  

$8.5-9.0$	&	$42$	&  0.16 $\pm$ 0.03	&  2.39 $\pm$ 0.55 &  11.2 $\pm$  4.2 &  0.23 $\pm$  0.08 &  3.29 $\pm$  0.55 \\  

$9.0-9.5$	&	$44$	&  0.35 $\pm$ 0.08	 & 9.25 $\pm$ 1.83  & 	41.2 $\pm$  11.0   & 0.34 $\pm$  0.04 &  8.82 $\pm$ \ 1.25   \\

$9.5-10.0$ &	$34$	&  0.48 $\pm$ 0.07	 &  14.7 $\pm$  2.5 	& 	 48.7	$\pm$ 7.0    &  0.41 $\pm$  0.03 &  12.9 $\pm$ 2.0\\  

$10.0-10.5$	&	$35$	&  0.64 $\pm$ 0.11	 &  9.08 $\pm$  1.87 	& 	32.6 $\pm$  6.4   &  0.42 $\pm$  0.05   &  6.74 $\pm$  1.22  \\ 

$10.5 -11.0$	&	$10$	&  0.38 $\pm$ 0.09	 &  0.58 $\pm$  0.13 	&  2.54 $\pm$  0.56   &  0.65 $\pm$  0.00  & 1.30 $\pm$ 0.63 \\  
	 & &  \\
	 With J0242+00 \\
	($9.0-9.5$)	&	(45)	&  (0.92 $\pm$ 0.40)	 & (24.3 $\pm$ 15.3)  & 	(64.7 $\pm$ 25.8)  &  (0.49 $\pm$ 0.04) &  (13.8 $\pm$ 5.1)   \\    
  \hline
		 & &  \\
\end{tabular}}

\caption{Contributions to SFE$_{\textsc{Hi}}$, SFRD and luminosity densities analysed by \textsc{Hi} mass bin.  Column descriptions [units]: Col. (1):  Log \textsc{Hi} mass range. Col. (2): Number of  \textsc{Hi}  targets within the \textsc{Hi} mass range.  Col. (3): SFE$_{\textsc{Hi}}$ value derived using \HA{} observations [\MS yr$^{-1}$]. Col. (4): SFRD contribution per  \textsc{Hi} mass bin [\MS yr$^{-1}$Mpc$^{-3}$dex$^{-1}$].  Col. (5) \textit{R}-band density contribution per decade of \textsc{Hi} mass [ergs s$^{-1}$\AA$^{-1}$Mpc$^{-3}$ dex$^{-1}$].  Col. (6): SFE$_{\textsc{Hi}}$ value derived using FUV observations [\MS yr$^{-1}$]. Col. (7) SFRD contribution per \textsc{Hi} mass bin [\MS yr$^{-1}$Mpc$^{-3}$dex$^{-1}$].  The $9.0 - 9.5$ bin appears twice; the lower listing (in brackets) shows the impact of J0242+00 on this mass bin, if it was included in the final sample (see discussion in Appendix \ref{sect:rem_gal_section}). The remaining mass bins are not shown as the luminosity densities do not change if J0242+00 was included, although their uncertainities would change, reflecting increased uncertainty from sampling. \label{table:sfrds_and_sfes_table}}    

\end{table*}

\section{Discussion}
\label{sec:Discussion_sec} 

\subsection{The local star formation rate density}   
\label{subsection:SFRDz0_sec}

The $ {\textnormal{\HA}}$ and FUV SFRD results 
are only marginally different (0.03 dex).  The similarity of the results from two distinct tracers occurs despite the strong systematic trends in the F$_{\text{\HA}}/f_{\text{FUV}}$ ratios outlined in Section \ref{sec:Results_sec}.  
The lowest \textsc{Hi} mass bin has a low \HA/FUV fractional luminosity density ratio of 0.63 (see Table \ref{table:ratios_full_table}$a$), with more central bins having higher values, up to \textit{l}$_{\textnormal{H}\alpha}$/\textit{l}$_{\textnormal{FUV}}$ of 1.26.  Scaling luminosities to the HIMF increases the \HA{} contributions sufficiently overall to offset the impact of the low \HA{} emission from low mass, low luminosity and LSB galaxies, and produces the near-identical \HA{} and FUV SFRD values reported here. 


Figure \ref{figure:forkai }$a$ shows that the SFRD results are towards the high end of the distribution of earlier z $\sim$ 0 measurements, including those  summarised in \citet {RN439}, \citet{RN81} and \citet{RN328}, and are consistent with the recent results of \citet{RN581}.  The SFRD values are also consistent, within errors, with another recent \textsc{Hi}-selected survey, \citet{RN483}, as well as with the first data release of 110 SINGG galaxies \citep{RN18}.

\begin{figure*}
\centering
\includegraphics[scale=0.84]{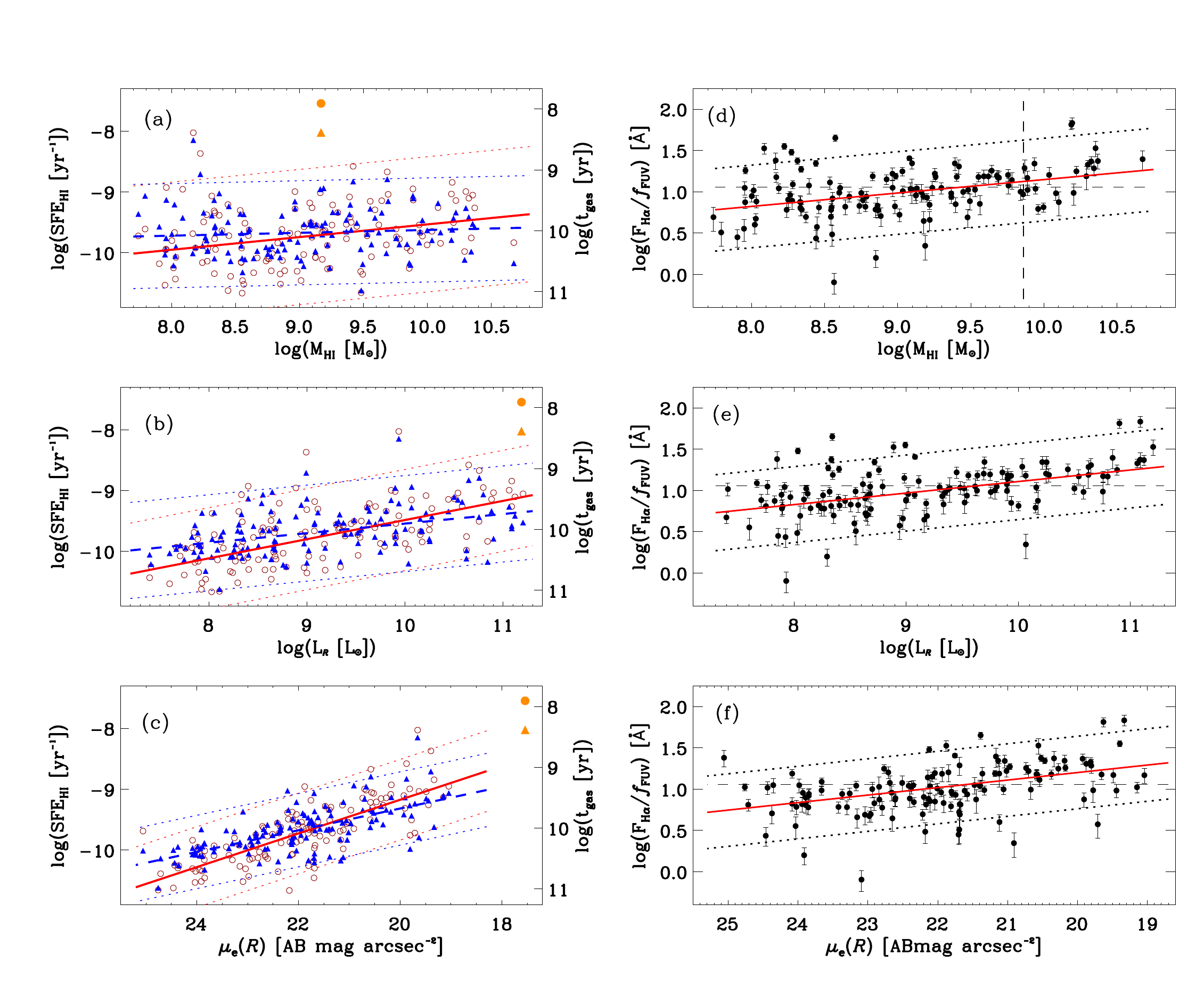}
\caption{Analysis of the 129 single galaxies in the sample with a signal-to-noise ratio (S/N) of over 3 for both \HA{} and FUV fluxes.  There are 160 single galaxies in the sample;  the 31 single galaxies not meeting the S/N requirement are not included in the analysis above. All quantities have been corrected for Galactic and internal dust absorption.  Panels (a) $-$ (c) \textsc{Hi}-based star formation efficiency (SFE$_{\textnormal{\textsc{Hi}}}$) as a function of key galaxy parameters.  SFE$_{\textnormal{\textsc{Hi}}}$(\HA{}) ($= $~SFR$_{\textnormal{H}\alpha}/$M$_{\text{\textsc{Hi}}})$ and  SFE$_{\textnormal{\textsc{Hi}}}$(FUV) (= ~SFR$_{\textnormal{FUV}}/$M$_{\text{\textsc{Hi}}}$) values are represented by red open circles and  blue filled triangles, respectively.  Solid red lines and dashed blue lines show the ordinary least squares best fit lines (Y vs. X, with a 2.5$\sigma$ iterative clipping) for \HA{} and FUV data, respectively.  See Table \ref{table:bestfits_table} for further details.   Dotted lines indicate $\pm$2.5$\sigma$ offsets to the fit, where $\sigma$ is the dispersion in the residuals of SFE$_{\textnormal{\textsc{Hi}}}$.  The uncertainties for individual galaxy \HA{} and FUV SFE values are smaller than the symbols used and and are not shown here.  J0242+00 SFE$_{\textnormal{\textsc{Hi}}}$(\HA{}) and  SFE$_{\textnormal{\textsc{Hi}}}$(FUV) values are overlaid with large filled orange symbols: circle and triangle, respectively.  J0242+00 has not been included in the determination of the best fit lines, or in other calculations (see Appendix \ref{sect:rem_gal_section} for further discussion on this galaxy). Panels (d) $-$ (f): Ratio of \HA{} line flux to FUV flux density as a function of \textsc{Hi} mass, \textit{R}-band luminosity and \textit{R}-band surface brightness; this ratio is equivalent to SFE$_{\textnormal{\textsc{Hi}}}$(\HA{})/SFE$_{\textnormal{\textsc{Hi}}}$(FUV).  Solid lines show the ordinary least squares best fit lines (Y vs X, with 2.5$\sigma$ iterative clipping);  see also Table \ref{table:bestfits_table}.  The horizontal dashed lines (panels d $-$ f) represent the expected F$_{\text{\HA}}/f_{\text{FUV}}$ value assuming a \citet{RN317} IMF using Eqn. 3 from \citet{RN322}.  The vertical dashed line in panel (d) shows the Schechter fit characteristic \textsc{Hi} mass (log(M$_{*}$/\MS) = 9.86)  of the \citet{RN318} HIMF. 
\label{fig:sixplot }
}
\end{figure*}

\begin{figure*}
\centering
\includegraphics[scale=0.70]{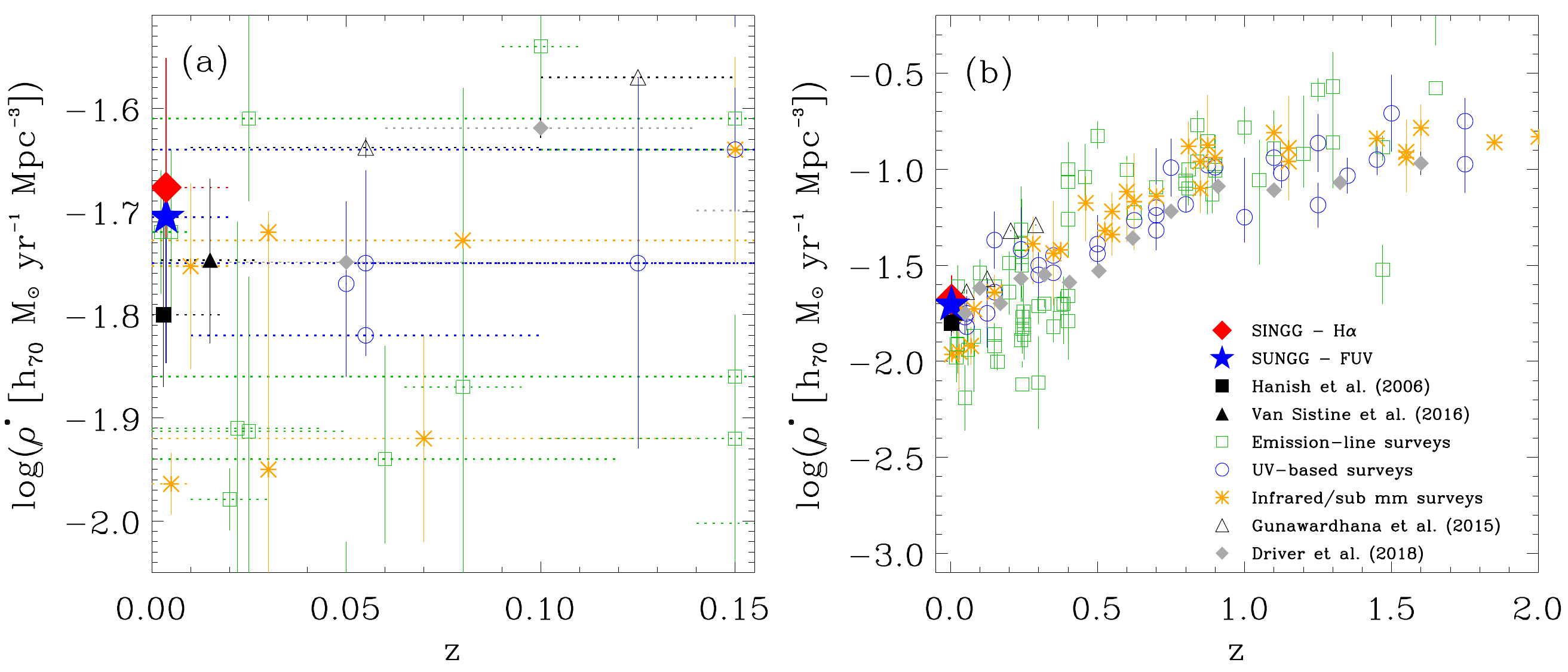}
\caption{SFRD, after correction for internal dust extinction ($\dot{\rho}$), as a function of redshift: (a) for z < 0.15 and (b) z < 2.0.  The SINGG  \HA{} result is marked with a large red diamond and the SUNGG FUV result with a large blue star.  The \textsc{Hi}-selected surveys, \citet{RN18} and \citet{RN483}, are shown with a filled black square and triangle, respectively.  Other values are sourced from compilations of results (\citet{RN439}, \citet{RN438},  \citet{RN81}) and recent research by \citet{RN581, RN328}, \citet{RN16} and \citet{RN687}.   Emission-line surveys (typically \HA), UV-based surveys and  infrared and submillimeter surveys are indicated with open green squares,  blue circles and orange asterisks, respectively.   The original $\dot{\rho}$ values have been adjusted, where necessary, to a uniform $\Lambda$CDM cosmology and a \citet{RN317} IMF, using a Hubble constant of H$_{0} = 70 ~$km s$^{-1}$ Mpc$^{-1},  ~\Omega_{0} =0.3$ and $\Omega_{\Lambda} =0.70$.  When dust correction information is not given in the source tables a 0.4 dex correction is assumed.  The vertical solid lines give the published uncertainty of each result and in (a) the horizontal dotted lines  indicate the redshift range applying to each sample.   
 \label{figure:forkai } }  
\end{figure*}

\subsection{F$_{\text{H$\alpha$}}/f_{\text{FUV}}$  variations}
\label{subsect:hafuvratio_sec}

F$_{\text{\HA}}/f_{\text{FUV}}$  varies systematically with several galaxy properties (see Figs. \ref{fig:sixplot }$d$ -- $f$), consistent with previous findings by \citet{RN322}  and others \citep[e.g.,][]{RN594}.  Undetected or unmeasured \HA{} emission would reduce the sample's F$_{\text{H$\alpha$}}/f_{\text{FUV}}$ ratio and \textit{l}$_{\textnormal{H}\alpha}$ contributions.  Detailed reviews of the observations ensured all discernible \HA{} flux was measured (see Section  \ref{subsection:SINGG_sec}).  \citet{RN415} used deep \HA{} observations in their work on dwarf galaxies, identifying previously undetected extended LSB \HA{} emission and determined an extrapolated effect of $\sim$5 per cent, insufficient to explain all of the low F$_{\text{H$\alpha$}}/f_{\text{FUV}}$ ratios in their research, or in our results. 

\citet{RN322} examined possible explanations for the F$_{\text{\HA}}/f_{\text{FUV}}$ variations.  Dust corrections and metallicity considerations are largely discounted as possible causes, with escaping ionising flux unable to be ruled out, while both stochasticity and a non-universal IMF are seen as plausible explanations.  Stochastic effects, due to the limited number of massive stars and short-lived intense star-forming periods, can account for some, if not all, of the observed IMF variations according to some recent research \citep [e.g.,][] {RN648,RN443, RN597,RN437, RN441, RN482, RN461}.  Lower mass galaxies ($\sim10^7 - 10^8$ \MS  ~particularly) may experience more intense episodes of star formation on shorter time-scales than other galaxies \citep[e.g.,][]{RN592, RN596, RN332}, so stochastic effects may be important in explaining at least some of the observed F$_{\text{\HA}}/f_{\text{FUV}}$ variations.

Stochastic effects aside, there is some theoretical support for IMF variations \citep[e.g.,][]{RN434,RN435,RN447} and growing observational evidence since \citet{RN322} that the IMF can vary with local conditions.  Variations in the low-mass end of the IMF have been observed in old galaxies not currently forming stars \citep[e.g.,][]{RN467, RN2} and in ultra-faint dwarf galaxies \citep[][]{RN600}, for example.  The upper end of the IMF may be suppressed due to local conditions in disk galaxies, with reduced massive star formation theorised or observed in low mass and low luminosity galaxies \citep[e.g.,][]{RN311}, in the less dense, outer regions of galaxies \citep[][]{RN485,RN432,RN1309} and also in LSB galaxies \citep[e.g.,][]{RN36,RN322}.   Top-light IMFs have also recently been inferred in galaxies with low star formation rates \citep[e.g.,][]{RN34,RN436}, in the centre of the Milky Way \citep{RN599} and where gas surface densities lie below the \citet[][]{RN433} critical density  \citep{RN485}.  Some recent studies suggest that the observations of  apparent IMF variations could be within the limits of statistical uncertainties, or are due to the flaws in the approach followed  \citep[e.g.,][]{RN6, RN598}.

\subsection{Systematic and random errors}
\label{subsection:Uncertainties_sec}

Table \ref{table:errors_himfsimple_table} lists the quantified random and systematic uncertainties in our luminosity density calculations.  Errors are generally calculated in accordance with the first data release \citep[for details see][]{RN18} and are  dominated by corrections for internal dust attenuation, HIMF model and sampling uncertainties.

\subsubsection{\textsc{HI} mass function selection}
\label{subsection:HIMFdiscussion_sec}

A key source of systematic error in the results is the HIMF used.  Recent studies have found evidence that the density of the environment affects the HIMF \citep[e.g.,][]{RN318, RN46, RN53, RN26}, so the HIMF selection requires careful consideration.   High density regions can exhibit a steeper HIMF slope at the low-mass end \citep[e.g.,][]{RN318,RN258}, although there are a number of contradictory results using different methodologies \citep[e.g.][]{RN51,RN26}.  The SINGG sample contains many loose groups  but few galaxies in clusters, consistent with findings that \textsc{Hi}-selected galaxies are less clustered than optically-selected samples with comparable luminosities \citep{RN606, RN603, RN604}.    

To estimate the impact of the HIMF selection, five  alternative published HIMFs were applied to the sample, keeping all other variables unchanged.   This approach also allows us to estimate the uncertainties due to  cosmic variance, as described in Section \ref{subsubsection{Cosmic variance}}.  Table \ref{table:HIMF_table} sets out the HIMFs and resultant SFRD and $\rho_{\textnormal{\textsc{Hi}}}$ values.  The HIMFs listed are derived from a variety of recent large volume surveys in \textsc{Hi}  \citep{RN318, RN51, RN39,RN22, RN678}.  The SFRD values derived vary  by up to 0.10 dex compared to our adopted HIMF model, reflecting the small differences in the individual HIMF parameters for these wide-field surveys (see Table \ref{table:HIMF_table}).  

\subsubsection{Cosmic variance}
\label{subsubsection{Cosmic variance}}
Due to the wide variety of galactic environments in the Universe, cosmic variance is a key source of uncertainty in all SFRD calculations \citep[e.g., see][]{RN641,RN581}.  By using HIPASS, a wide-field \textsc{Hi} survey, and sampling the entire \textsc{Hi} mass range, SINGG/SUNGG reduces the sampling biases that can become significant in surveys with smaller sampling volumes.   The working assumption is that the mix of galaxy types depends only on \textsc{Hi} mass and is well represented by our sample.

By design the SINGG and SUNGG surveys are not volume-complete.   Galaxies were instead chosen to fully sample the HIMF and, within individual mass bins, the nearest galaxies were preferentially selected to optimise spatial resolution \citep[see][]{RN41}.  

The impact of cosmic variance can then be assessed by comparing SFRD values derived from using HIMFs taken from different wide-field surveys (see Section \ref{subsection:HIMFdiscussion_sec}) and, in particular, by using HIMFs from survey volumes with significantly different environmental characteristics.  Applying the ALFALFA Survey's \citep{RN678} Spring HIMF (overdense and Virgo Cluster-dominated) and the Fall HIMF (underdense and void-dominated), for example, generates \HA{} SFRD values for our sample of 0.0248 and 0.0189 [M$_{\odot} $ yr$^{-1} $ Mpc$^{-3}]$, respectively.  The $\sim$ 0.12 dex difference in the SFRD values is similar to the uncertainties arising from all other random and systematic sources, highlighting the importance of cosmic variance in the error analysis.   

Using increasingly larger volume surveys for measuring the local SFRD can reduce cosmic variance uncertainties.  Due to flux-detection limits, however, the accessible volume for low luminosity and LSB galaxies remains constrained by observational capabilities. With low mass (e.g., log(M$_{\text{\textsc{Hi}}}/$M$_\odot) < 9.0$) and low luminosity galaxies contributing over 20  per cent of local \HA{} and FUV SFRD values (see Table \ref{table:ratios_full_table}), this is a significant constraint on the completeness of SFRD measurements.

\subsubsection{Distance model}
\label{subsubsection{Distance_sec}}

To gauge the systematic uncertainty arising from our choice of distance model, the SFRD was recalculated using the local-group distances of \citet{RN318}.  This increases $ \dot{\rho}_{{{\textnormal{H}\alpha}}}$ and $\dot{\rho}_{{\textnormal{FUV}}}$ by 0.022 dex and 0.011 dex, respectively.  These values have been taken as the systematic error arising from the distance model selected (see Table \ref{table:errors_himfsimple_table}). 

\subsubsection{[\textsc{Nii}] \textit{contamination and  internal dust attenuation} }

The empirical relationship between [\textsc{Nii}] fluxes and uncorrected \textit{R}-band magnitudes of \citet{RN584}, derived from  The Nearby Field Galaxy Survey \citep{RN653},  is used to adjust SINGG \HA{} fluxes for both internal dust attenuation and [\textsc{Nii}] contamination.  \citet{RN620} also uses this consistent approach, but most surveys have attenuation and [\textsc{Nii}] corrections derived from different galaxy populations.  Commonly used alternatives for the [\textsc{Nii}] corrections apply the empirical relationships of   \citet{RN593,RN652,RN614} or simplistically reduce  \HA{} fluxes by a fixed value, often based on one or more of these references.  \citet{RN653} showed, however, that the [\textsc{Nii}]/\HA{} flux ratio was more closely related to galaxy luminosity than morphology, and that earlier empirical relationships consistently over-correct for galaxy-wide [\textsc{Nii}] contamination.   \HA{} fluxes are adjusted by a factor of 0.05 ($-0.12$ AB mag) for [\textsc{Nii}] contamination and  \HA{} and FUV fluxes are adjusted by factors of $-0.26$ and $-0.38$ ($+0.66$ and $+0.96$ AB mag), respectively, for dust attenuation.

\subsubsection{Stellar absorption and other errors}

\citet{RN645} determined \HA{} stellar absorption corrections ranging from 2 $- $ 6 per cent were needed to the measured \HA{} fluxes and the mid-range of these values (4  per cent) is used to increase SINGG \HA{} and EW(\HA) measurements.  Recent research shows average stellar absorption can vary systematically with galaxy luminosity \citep[e.g.,][]{RN656} and galaxy mass \citep[e.g.,][]{RN655}, leading to an underestimation of the SFRD \citep[see also][]{RN658}.  Due to the relatively small contribution the stellar absorption correction makes to the total uncertainty (see Table \ref{table:errors_himfsimple_table}) we do not apply a more elaborate correction.

\section{Conclusions}
\label{sec:Conclusions_sec} 
 
We have presented the first parallel \HA{} and FUV-derived star formation rate density values obtained from an \textsc{Hi}-selected sample of nearby galaxies.    
We find a consistent SFRD of $\sim$ 0.020 [M$_{\odot} $ yr$^{-1} $ Mpc$^{-3}]$ for the two measurements, with a difference between the two measurements which is within the 1-$\sigma$ uncertainties of each} ($\sim$ 0.13 dex).  Figure \ref{figure:forkai } shows these measurements lie towards the top of the distribution of recent results, reflecting the more complete nature of our \textsc{Hi}-selected sample, which is less biased against low luminosity and low surface brightness galaxies.

The HIMF-based methodology has been used by \citet{RN18} and \citet{RN483} and our results are consistent with theirs.  This method facilitates the efficient derivation of SFRD and other volume densities, particularly when observing resources  are limited.  The thorough sampling along the HIMF, which forms the foundation for the sample selection, also leads to relatively better testing of the low \textsc{Hi}-mass regime, compared to  most optically-selected samples.  The approach is supported by recent comparisons with the more commonly applied $V_{\textnormal{max}}$-based correction in volume-incomplete samples \citep[see e.g.,][]{RN589, RN590, RN483}, but is susceptible to extreme outliers, as experienced here with J0242+00.   

The similarity of SFRD from the two SFR indicators occurs despite significant differences in the F$_{\text{\HA}}/f_{\text{FUV}}$ values in the sample.    Galaxies with lower surface brightness, luminosity or \textsc{Hi} mass, tend to have lower F$_{\textnormal{H}\alpha}/f_{\textnormal{FUV}}$ values than those at the high end of those parameters.  This ratio is equal to what is expected for a Salpter IMF for galaxies near M$_{\text{\textsc{Hi}}}^{*}$; the fiducial \textsc{Hi} mass in the Schechter mass function fit.  The trends suggest IMF variations may be in effect at the extreme ends of this parameter space.    

\section*{Acknowledgements}

We thank the anonymous referee for constructive and detailed comments that have improved this paper.  Partial funding for the SINGG and SUNGG surveys came from NASA grants NAG5-13083 (LTSA program), GALEX GI04-0105-0009 (NASA GALEX Guest Investigator grant) and NNX09AF85G (GALEX archival grant) to G.R. Meurer.  FAR acknowledges partial funding from the Department of Physics, University of Western Australia.  This research has made use of the NASA/IPAC Extragalactic Database
(NED), which is operated by the Jet Propulsion Laboratory, California
Institute of Technology, under contract with the National
Aeronautics and Space Administration.



\bibliographystyle{mnras}
 \bibliography{singgbib}


\appendix

\section{A Remarkable Galaxy}
 \label{sect:rem_gal_section}
 
\begin{figure}
\centering
\includegraphics[scale=0.44]{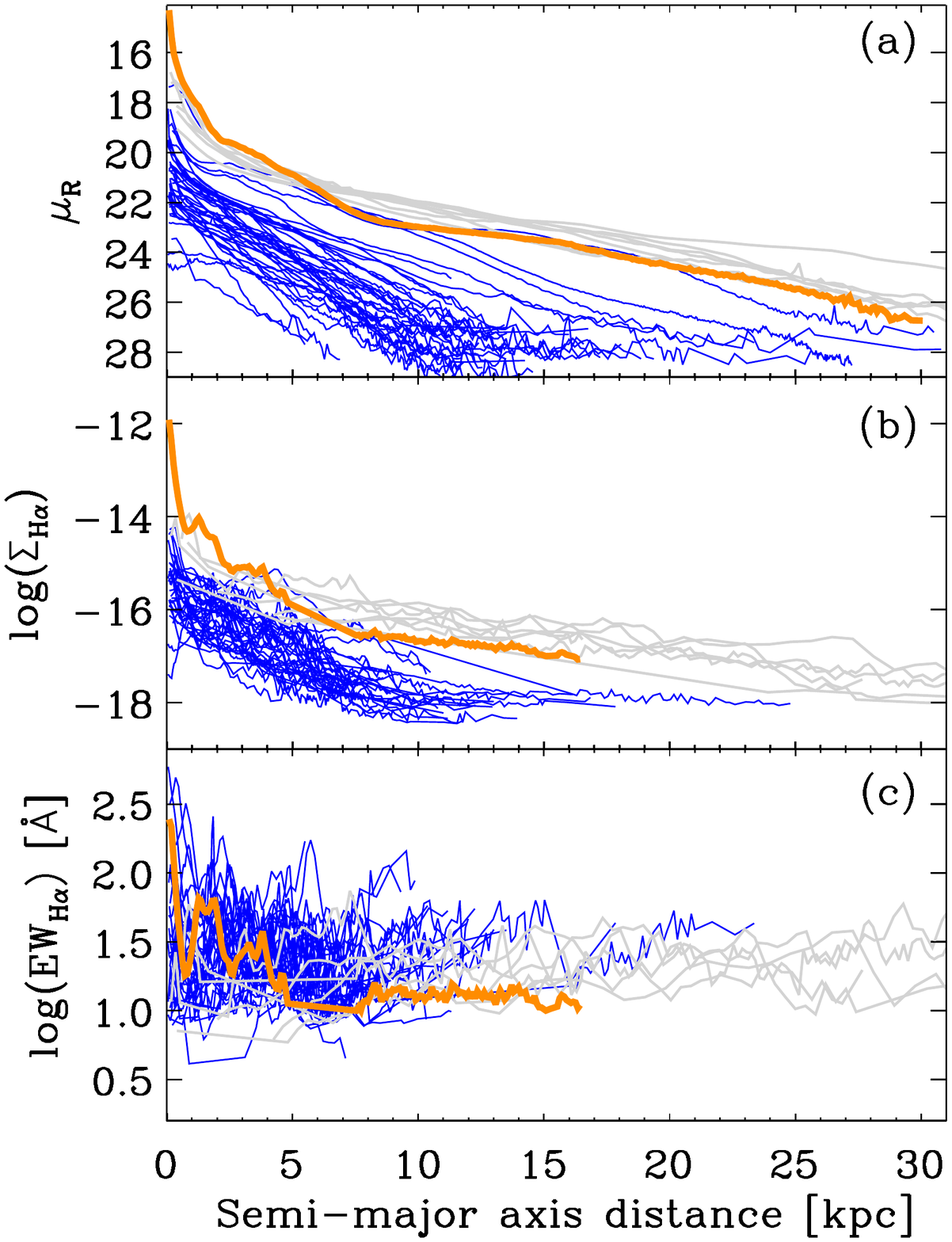}
\caption{The radial surface brightness profile of J0242+00 (shown with a thick orange line) dominates within the inner $\sim$ 3~kpc, in comparison to the other seven  highest luminosity ($M_{R} < -23$ AB mag) single galaxies in the sample (profiles shown in light grey) and the 40 other single galaxies in the same \textsc{Hi} mass bin (i.e., log(M$_{\text{\textsc{Hi}}}$/\MS) $= 9.0 - 9.5$, shown with thin blue lines).  Data series with less than 5 points, or where S/N $<3.0$, are excluded.  Panel (a) shows the $R$-band radial surface brightness profiles  in AB mag arcsec$^{-2}$ and panel (b) shows the  log of the \HA{} surface brightness in units of erg cm$^{-2}$ s$^{-1}$ arcsec${^{-2}}$.  The profiles are adjusted to face-on values (i.e., raw intensities multiplied by the minor to major axis ratio (b/a) of the elliptical apertures used to extract the profiles).  Panel (c) shows the equivalent width   [\AA] derived from the ratio of \HA{} and $R$-band intensities.
\label{fig:J0242plots }}

\end{figure}

The nearby (D $\sim$ 16.2 Mpc) galaxy HIPASS J0242+00, better known as NGC 1068, would contribute a phenomenal 27\%,  12\% and 14\% of the total cosmic luminosity densities in \HA{}, FUV and \textit{R}-band, respectively, derived using our methodology, if it was included in the sample (see Table \ref{table:outliers_table}).  This reflects its remarkable luminosity, especially for its \textsc{Hi} mass (M$_{\text{\textsc{Hi}}} = \sim 10^{9.2}$ \MS) and is largely a by-product of our HIMF-based methodology.  In a volume-complete sample J0242+00  probably would not have such an impact, however.

Figure  \ref{fig:J0242plots } shows this archetypal Type II Seyfert galaxy \citep{RN49} has extraordinarily intense emission, especially compared to galaxies having a similar \textsc{Hi} mass, but also compared to galaxies of similar luminosity for radii less than $\sim 3$ kpc.   It is one of the most luminous objects known in the local Universe \citep[e.g., see][]{RN8} and only one of eight galaxies in our sample with M$_{R}$ < -23 AB mag.  The central region (r < 2.3"/180 pc) contributes 5 and 30 per cent of the galaxy's total \textit{R}-band and \HA{}  fluxes respectively.  Intense star formation is occurring within this small radius \citep{RN66,RN426} and, therefore, the AGN makes a minor direct contribution to the galaxy's total \textit{R}-band and \HA{} luminosities ($6.64~\times  10^{40}$ ergs s$^{-1} \angstrom^{-1}$ and $ 4.36~\times 10^{42}$ ergs s$^{-1},$ respectively).  Similarly, \citet{RN334} found that most of the galaxy's FUV flux does not originate from the AGN, but instead is predominately ($\sim81\%$)  generated in the galaxy's disk. 

The unusually high surface brightness disk contains star forming knots of extraordinary mass and luminosity \citep[see][]{RN316,RN8,RN423}.  These knots occur out to $\sim3$~kpc from the central AGN region \citep{RN429} and cause the rises in the radial profiles illustrated in Figure \ref{fig:J0242plots }\textit{b} and \textit{c}  \citep[see also][]{RN316,RN313}.  This intense star formation, just outside the nucleus, is thought to arise from bar-driven gas flows, rather than being AGN-driven \citep[see][]{RN307,RN417,RN72, RN423}.  

The disproportionate impact of J0242+00, if it were included in the final sample, partly reflects the small size of the SINGG and SUNGG surveys.   It has therefore been excluded from our analysis and results.


\section{Significant HIPASS Targets}
\label{sect:sig_HIPASS}

\begin{table*}
		\normalsize{Galaxies with the largest impact on  $\textit{l}_{\textnormal{H}\alpha}$ and \textit{l}}$_{\textnormal{FUV}}$ \\
	\centering
\resizebox{\linewidth}{!}{%
	\begin{tabular}{llcllllll} 
		
		\hline
		 \textsc{Hi} target         & {   log}      & \textit{l}$_{\textnormal{H}\alpha}$   &  \textit{l}$_{\textnormal{FUV}}$    &   Notes\\  
		& (M$_{\text{\textsc{Hi}}}$/\MS)&  fraction  & fraction &  \\
		  (1)&  &  (2)&  (3) &    \\   
		\hline
	     & & &\\
	     
	      J1338-17 	& $~~9.69$  & 0.046 & 0.039  & NGC 5247: grand-design spiral \citep{RN610}\\ 
	      J1247-03 & $~~8.17$  	& 0.043 & 0.036  & NGC 4691: central starburst and outflows \\ 
	      &&&& \citep[][]{RN609,RN635} \\
	      J0505-37 	& $~~9.42$  & 0.041 & 0.026  & NGC 1792: interacting with J0507-37 (below)\\ 
	      J1059-09   & $10.05$ & 0.040 & 0.027 & Group: 10 galaxies with \HA{} observations, 9 FUV \\ 
	      J0342-13  & $~~9.86$ & 0.028 & 0.017  & NGC 1421 Group:   2 galaxies with \HA{} observations, 1 FUV. \\ 
 		 J0216-11c  & $~~9.96$ & 0.026 & 0.024 &  NGC 873 \\  
          J0507-37   & $~~9.53$ & 0.024  & 0.027 & NGC 1808: interacting with J0505-37 (above)\\ 
\hline
		\hline

	\end{tabular}}
	\caption{Col. (1) The HIPASS targets with the largest impact on  \HA{} and FUV luminosity densities ($\textit{l}_{\textnormal{H}\alpha}$ and \textit{l}$_{\textnormal{FUV}}$, respectively).  Cols. (2 -- 3) The fraction of \textit{l}$_{\textnormal{H}\alpha}$ and \textit{l}$_{\textnormal{FUV}}$ arising from the listed targets.  In comparison, if included, J0242+00 (log(M$_{\text{\textsc{Hi}}}$/\MS) = 9.17) would make an extraordinary fractional contribution of 0.269 and 0.116  of increased  \textit{l}$_{\textnormal{H}\alpha}$ and \textit{l}$_{\textnormal{FUV}}$ values, respectively.
\label{table:outliers_table}}
\end{table*}


\section{Best Fit Lines}
\label{sect:bestfitable}

\begin{table*}
	\centering
\resizebox{0.9\linewidth}{!}{%
  
\begin{tabular}{lllllllll} %
&& \\
	Best fit line coefficients  \\    
		\hline
 Figure description	&	  Fig. ref. 	&	 	Flux & A 	&	B	 & $\sigma_{x}$ &$\sigma_{y}$ & N\\
 & & & (1)& (2)& (3) & (4)& (5)\\
 \hline
& &\\
SFE v log(M$_{\text{\textsc{Hi}}}/$M$_\odot$) & \ref{fig:sixplot }$a$ & \HA &$-11.62 \pm 0.49$ & $~~0.21 \pm 0.05$ & $2.13$ & $0.44$ &  $127$ \\
& & FUV & $-10.09 \pm 0.39$  & $~~0.05 \pm 0.04$ & $7.28$ &  $0.34$ & $127$  \\

SFE v log(L$_{\text{R}}$) & \ref{fig:sixplot }$b$ &  \HA & $-12.63 \pm 0.27$ & $~~0.32 \pm 0.03$ & 1.06 & 0.33  & 124\\

& & FUV & $-11.12 \pm 0.26$ & $~~0.16 \pm 0.03$ & $2.00$& $0.32$&  124\\

SFE v log($\mu_{R}$) & \ref{fig:sixplot }$c$ &  \HA & $~-3.64 \pm 0.38$ & $-0.28 \pm 0.02$  & $0.96$ & $0.27$  & $124$ \\

& & FUV & $~-5.75 \pm 0.32$ & $-0.18 \pm 0.01$ & $ 1.35$ & $0.24$  & 124\\

& &\\

\\
log(F$_{\textnormal{H}\alpha} /f_{\textnormal{FUV}}$) v log(M$_{\text{\textsc{Hi}}}/$M$_\odot$) & \ref{fig:sixplot }$d$ &  & $~-0.48 \pm 0.25$ & $~~0.16 \pm 0.03$ & 1.23 & 0.20  & $119$\\
 
log(F$_{\textnormal{H}\alpha} /f_{\textnormal{FUV}}$) v log(L$_{\text{R}}$) & \ref{fig:sixplot }$e$ &  & $~-0.29 \pm 0.16$ & $~~0.14 \pm 0.02$ & $1.32$ & $0.18$  & $118$ \\

log(F$_{\textnormal{H}\alpha} /f_{\textnormal{FUV}}$) v log($\mu_{R}$) & \ref{fig:sixplot }$f$ &  & $~~~~~3.02 \pm 0.27$ & $-0.09\pm 0.01$ & $ 1.93$ & $0.18$ & $115$ \\

& & \\

  \hline
		
	\end{tabular}}
		\caption{Coefficients and residuals of the best fit lines (ordinary least squares Y vs. X, using a 2.5$\sigma$ cut) in Fig. \ref{fig:sixplot }.  Column descriptions:   Cols. (1,2) coefficients of the best fit line, where y $= $ A + Bx, together with their 1$\sigma$ standard deviation values.  Cols. (3,4) x and y residual dispersions, respectively.  Col. (5) Number of galaxies used in the final fit, after iterative clipping (from a total population of 129 single galaxies meeting the S/N requirements described in Fig. \ref{fig:sixplot }).  \label{table:bestfits_table} }

\end{table*}

\section{ERROR ANALYSIS}
\label{sect:error_analysis}

\begin{table*}
	\centering
	
 \resizebox{0.85 \linewidth}{!}{%
 
\begin{tabular}{lllllllllll} %
	Error analysis  of log(luminosity densities)\\    
		\hline
 Uncertainties of log(luminosity density):	&	    	 &\textit{l}$_ {R}$& \textit{l}$_ {R}$ & \textit{l}$_{\textnormal{H}\alpha}$ 	&	\textit{l}$_{\textnormal{H}\alpha}$  & \textit{l}$_{\textnormal{FUV}}$  &  \textit{l}$_{\textnormal{FUV}}$  \\
   &&&(dust- && (dust- && (dust-\\
  &Notes& (uncorrected)&corrected)&(uncorrected)& corrected) & (uncorrected) & corrected)\\
 \hline
\textit{Random errors}  & \\
Sampling  & $\scriptstyle (1)$ &  $\scriptstyle ^{+0.040}_{-0.042}$  & $\scriptstyle ^{+0.044}_{-0.047}$ & $\scriptstyle^{+0.037}_{-0.043}$ & $\scriptstyle^{+0.038}_{-0.047}$ & $\scriptstyle^{+0.040}_{-0.043}$ & $^{+0.035}_{-0.037}$\\[0.075cm]
Sky subtraction  & $\scriptstyle (2)$& $\scriptstyle \pm{0.001}$ &$\scriptstyle \pm{0.001}$&$\scriptstyle \pm{0.002}$  & $\scriptstyle \pm{0.002} $ & $\scriptstyle^{+0.020}_{-0.022}$ & $^{+0.065}_{-0.087}$\\ [0.075cm]

Continuum subtraction & $\scriptstyle (3)$&  ~~{        ... }  & ~~{        ... }  & $\scriptstyle^{+0.008}_{-0.009}$ & $\scriptstyle \pm{0.011}$ & ~~{        ... }  & ~~{        ... } \\ [0.075cm]
Flux calibration & $\scriptstyle (4)$ & $\scriptstyle \pm{0.008}$ &$\scriptstyle \pm{0.008}$ &{$\scriptstyle \pm{0.011}$} & {$\scriptstyle \pm{0.010}$} & $\scriptstyle \pm{0.047}$ & $\scriptstyle \pm{0.047}$ \\ [0.075cm]
\textsc{[Nii]} correction & $\scriptstyle (5)   $&~~{        ... }  &  ~~{        ... } & $\scriptstyle^{+0.003}_{-0.006}$ & $\scriptstyle^{+0.004}_{-0.007}$ & ~~{        ... }  & ~~{        ... } \\ [0.075cm]
Internal dust extinction  & $\scriptstyle  (6)$ & ~~{        ... } &$\scriptstyle^{+0.043}_{-0.007}$&   ~~{        ... } & $\scriptstyle^{+0.121}_{-0.007}$ & ~~{        ... } & $\scriptstyle^{+0.002}_{-0.018}$ \\ [0.075cm]

\cmidrule(r{.7cm}){3-3}\cmidrule(r{.7cm}){4-4} \cmidrule(r{.7cm}){5-5}\cmidrule(r{.7cm}){6-6}\cmidrule(r{.7cm}){7-7}\cmidrule(r{.7cm}){8-8}

Total random errors   & & $\scriptstyle^{+0.041}_{-0.043}$ &$\scriptstyle^{+0.062}_{-0.048}$ &  $^{+0.040}_{-0.046}$ & $^{+0.128}_{-0.050}$ & $^{+0.065}_{-0.067}$ & $^{+0.090}_{-0.107}$ \\
 & & \\
\textit{Systematic errors} \\

\textsc{[Nii]} zero point & $\scriptstyle (7)$ & ~~{        ... } & ~~{        ... } & $\scriptstyle \pm{0.002}$ & $^{+0.003}_{-0.002}$ & ~~{        ... }  & ~~{        ... } \\[0.075cm]
Internal dust zero point & $\scriptstyle (8) $&  ~~{        ... } & $\scriptstyle \pm{0.003} $ &  ~~{        ... } & $\scriptstyle  \pm {0.006}$ &  ~~{        ... }  &  $^{+0.079}_{-0.078}$ \\ [0.075cm]

Distance model & $\scriptstyle (9) $&  $\scriptstyle +0.014 $& $\scriptstyle +0.018 $ &$\scriptstyle +0.017$ & $\scriptstyle +0.022$ & $\scriptstyle -0.005$ & $\scriptstyle +0.011$\\ 

\textsc{Hi} mass function & $\scriptstyle (10) $ & $\scriptstyle  \pm {0.004}$&$\scriptstyle  \pm {0.010}$&$\scriptstyle \pm{0.023}$ & $\scriptstyle \pm{0.008}$ &  $\scriptstyle \pm{0.037}$ &$\scriptstyle \pm{0.008}$ \\ [0.075cm]
\cmidrule(r{.7cm}){3-3}\cmidrule(r{.7cm}){4-4} \cmidrule(r{.7cm}){5-5}\cmidrule(r{.7cm}){6-6}\cmidrule(r{.7cm}){7-7}\cmidrule(r{.7cm}){8-8}

Total systematic errors & &   $\scriptstyle^{+0.015}_{-0.004}$& $\scriptstyle ^{+0.021}_{-0.010}$ &  $\scriptstyle ^{+0.029}_{-0.023}$ &  $\scriptstyle^{+0.024}_{-0.010}$ &  $\scriptstyle \pm{0.037}$  & $\scriptstyle ^{+0.080}_{-0.078}$   \\

 \cmidrule(r{.7cm}){3-3}\cmidrule(r{.7cm}){4-4} \cmidrule(r{.7cm}){5-5}\cmidrule(r{.7cm}){6-6}\cmidrule(r{.7cm}){7-7}\cmidrule(r{.7cm}){8-8}

Total  errors & &$\scriptstyle \pm{0.043}$ &$^{+0.065}_{-0.049}$&  $^{+0.049}_{-0.051}$ & $^{+0.130}_{-0.051}$ & $^{+0.075}_{-0.077}$ & $^{+0.120}_{-0.133}$ \\
& &\\
& & \\

  \hline
		
	\end{tabular}}
		\caption{Analysis of luminosity density uncertainties (log values), for uncorrected and dust-corrected R, \HA{} and FUV fluxes.  All errors, excluding (10),  have been calculated in accordance with \citet[][]{RN18}.  Notes: (1) The sampling error is the standard deviation of the results from bootstrapping 10,000 samples of 294 randomly selected galaxies (duplication permitted). (2 \& 3) Sky and continuum subtraction uncertainties  are the standard deviations from 10,000 iterations where sky level and continuum levels were randomly altered for each galaxy within the error model. (4) The \HA{} flux calibration uncertainty is estimated at 0.04 mag for images using the 6568/28 narrowband filter and 0.02 mag for all others. FUV flux calibration uncertainties are in accordance with  \citet{RN691,RN692}.  (5 \& 6)  The underlying $M_{R}'$ fits  of \citet[][private communication]{RN584} have a  0.23 dex dispersion arising from uncertainity in internal dust extinction and a  0.23 dex dispersion due to [\textsc{Nii}] correction.  The quoted random errors are the standard deviations from two separate 10,000 realisations where each galaxy's corrections were randomly altered with a 0.23 dex dispersion around the mean.  (7 \& 8)  The zero-point error associated with the $M_{R}'$ fits random uncertainties (see \citet{RN18}). (9) The quoted error is the difference in the SFRDs  derived using our default \citet{RN424}  model and the SFRD using the alternative local-group distances of \citet{RN318} (see Section \ref{subsubsection{Distance_sec}}). (10) The HIMF uncertainites are the differences in the derived SFRDs from using the default \citet{RN318} HIMF compared to  the average of the five alternative wide-field survey HIMFs listed in Table \ref{table:HIMF_table} (see Section \ref{subsection:HIMFdiscussion_sec}). \label{table:errors_himfsimple_table} }

\end{table*}


\bsp	
\label{lastpage}
\end{document}